\documentclass[a4paper,fleqn]{cas-dc}

\usepackage[authoryear,longnamesfirst]{natbib}

\usepackage{amssymb}
\usepackage{lipsum}
\usepackage{amsmath}
\usepackage{balance}
\usepackage[dvipsnames]{xcolor}
\usepackage{multicol}
\usepackage{natbib}
\setcitestyle{numbers,comma,square}
\usepackage{makecell}
\usepackage{url}
\usepackage[labelfont=bf,justification=raggedright,singlelinecheck=false]{caption}
\usepackage{soul}  

\def\tsc#1{\csdef{#1}{\textsc{\lowercase{#1}}\xspace}}
\tsc{WGM}
\tsc{QE}
\tsc{EP}
\tsc{PMS}
\tsc{BEC}
\tsc{DE}

\begin{document}
\let\WriteBookmarks\relax
\def\floatpagepagefraction{1}
\def\textpagefraction{.001}

\shorttitle{Tunable Coloration }
\shortauthors{Md. Shariful Islam et~al.}

\title [mode = title]{Tunable Coloration in Core-Shell Plasmonic Nanopixels Based on Organic Conductive Polymers: A First-Principles and FDTD Study}                      



%
\author[1]{Md. Shariful Islam}[type=editor,
                        auid=000,bioid=1,
                        orcid=0000-0001-7922-096X]



\ead[url]{sites.google.com/student.sust.edu/md-shariful-islam}

\credit{ Conceptualization, Methodology, Visualization, Software, Investigation, Writing - Original Draft.}

\author[1]{ Ahmed Zubair}[style=chinese, orcid=0000-0002-1833-2244]
\cormark[1]
\ead{ahmedzubair@eee.buet.ac.bd}
\ead[URL]{sites.google.com/view/ahmed-zubair-research-group/home}

\credit{Conceptualization, Methodology, Visualization, Resource, Supervision, Writing - Original Draft, Writing-Review \& Editing.}

\cortext[cor1]{Corresponding author}

\affiliation[1]{organization={Department of Electrical and Electronic Engineering, Bangladesh University of Engineering and Technology}, 
    city={Dhaka},
    postcode={1205}, 
    country={Bangladesh}}



\begin{abstract}
From raindrops to planets, the scattering of electromagnetic fields introduces exciting phenomena that can be utilized for display devices. Here, we designed an electrochromic nanoparticle on mirror (eNPoM) structure with core-shell geometries for low-power nanoscale pixels with rapid coloration abilities based on four electrochromic organic conducting polymers utilizing the first-principles calculations based on density functional theory (DFT) and the finite-difference time-domain (FDTD) simulations. Au nanoparticles are coated with electrochromic conductive polymers (such as PANI, PEDOT, PPy, and PTh) and positioned on the metal mirror. The electric field enhancement and the impact of shell thickness are analyzed. Dielectric properties of all polymers resulting from atomistic calculation were utilized for FDTD simulation, which helps to correlate the direct relationship between polymer structure and optical properties. Notably, the study reveals significant wavelength tunability of 100nm, 40nm, 70nm, and over 40nm using PANI, PEDOT, PPy, and PTh shells, respectively. Additionally, the potential for RGB color production using a TiN layer on the mirror is explored. For the first time, complex structures such as bow tie and gear were utilized to model the nanopixels studied and a significant absorption peak shift was observed. Chromaticity coordinates in the CIE 1931 color space and CIELAB2000 color difference quantify color change capabilities during the redox cycle, and a comparative analysis of organic and inorganic materials highlights the prospects of the proposed plasmonic nanopixels. A comparative analysis of organic versus inorganic materials highlights the advantages of the proposed plasmonic nanopixels. These electrochromic conducting polymer-based plasmonic nanopixels are expected to surpass current display technologies in color dynamics and refresh rates, with potential applications in diverse display systems.

\end{abstract}



\maketitle{}
\section{Introduction}
Electrochromic devices (ECDs) exhibit a remarkable ability to modulate their transparency and refractive index in response to external stimuli such as external electrical stimulation\,\cite{wang2020fusing}. Electrochromic materials have been extensively used in numerous applications such as optical filters, optical switches, bi-stable devices, energy-efficient smart windows, wearable electrochromic devices, and variable optical attenuators\,\cite{Qi_2003_EC_VOA}. This wide-range applicability can be attributed to their easy synthesis, less-effort fabrication, and higher stability compared to inorganic conducting materials. The next decade would be a boom for such versatile and energy-efficient devices, ranging their application from roadside billboards to smart goggles. Hence, research on nanocomposite organic/inorganic/hybrid electrochromic material and nanostructured plasmonic electrochromic devices has increased tremendously in recent decades.

\begin{figure*}[h]
\centering
\includegraphics[width=1\textwidth]{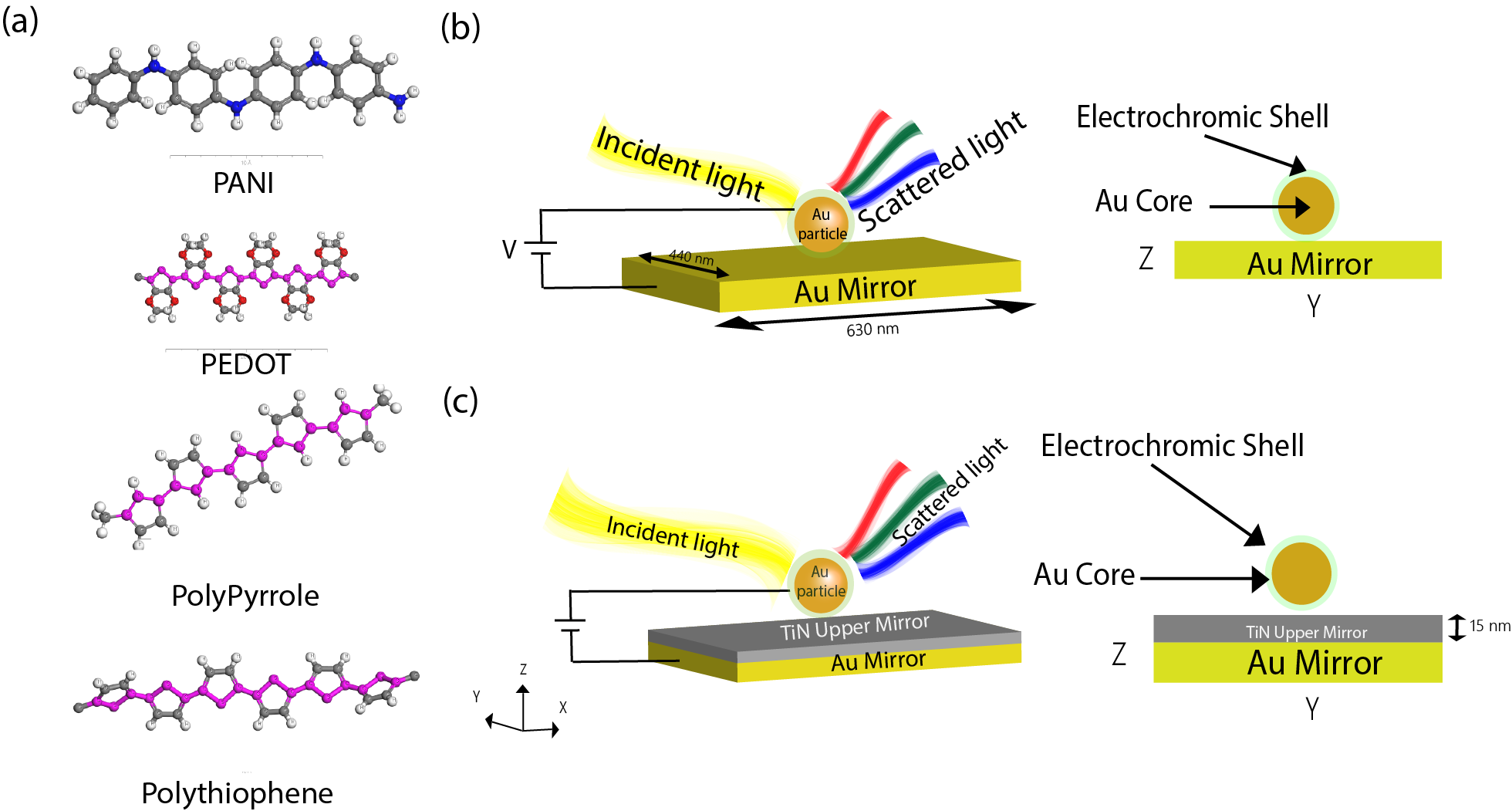}
\caption{(a) Polymer chain of selected organic electrochromic materials, (b) cross-sectional view of proposed nanopixel, and (c) illustrative structure of proposed NPoM structure with shell material composition.}
\label{structure}
\end{figure*}

Color refers to the perception of our eyes depending on the wavelength range of light which enters into. Choromism can be resulted from dyes pigments or from structural coloration. Sturctural coloration techniques are based on periodic grating such as photonic crystals or nanoparticle-field coupling i.e. plasmons. Structural coloration techniques are free from toxic compound involvement and have long-term durability.
One of the major requirements of modern optical devices, such as fast switchable plasmonic pixel-based displays\,\cite{kim_transparent_flexible_electrochromic_2019}, interactive panels,\cite{kim_2022_stretchable_electrochromic} and color filter\cite{soikot_metasurface_color}, is easy and fast switching and tuning. Till now, in the Red-Green-Blue (RGB) emission device industry, LEDs have the biggest share due to their smaller dimension and comparatively higher resolution than LCD. However, they lack flexibility, are not scalable, have much poorer performance in the presence of bright sunlight, have low quantum efficiency, contain toxic elements, and, most importantly, have a limited lifetime. Plasmonic nanopixels can be a possible solution for cheap, vivid, and flexible display requirements, as their strong light-matter interaction results in a unique, vibrant color generation. However, very few studies have been carried out on the design and optimization of such easily tunable devices. Most plasmonic devices are controlled by various stimuli such as radius, aspect ratio, particle orientation, symmetry, thickness, and fill fraction, which aren’t easy to modulate once manufactured. Hence, researchers are focusing on phase-changing materials, such as electrochromic materials,  having a dependency on the surrounding medium to switch plasmon resonant frequency\cite{peng_2019_nanopixel,peng_2022_electrochromic}. 
A chemical manifestation of electrochromism for various organic and inorganic compounds was analyzed by Rowley \textit{et al.} \cite{rowley2002chemistry}. Electron transport due to simultaneous insertion-extraction of ions is responsible for such behavior \,\cite{Ma2017_chemistry_behind_ec}. One notable characteristic of electrochromic devices is their remarkably low power requirement. Moreover, the transparency of electrochromic materials can be adjusted by applying significantly lower direct current potential, even as low as a fraction of volts\,\cite{thakur_2012_ec_volt}. The electrochromic materials have been reported for various applications such as flexible skin-attached stretchable interactive display,\cite{kim_transparent_flexible_electrochromic_2019,kim_2022_stretchable_electrochromic}, selective filter\,\cite{Anjali_2020_Polythiophene-nanoWO3-electrochromic-filter}.

Inorganic materials such as tungsten dioxide (WO$_2$) and other transitional metal oxide-based thin films are promising for high-speed electrochromic devices due to their transmittance modulation property\,\cite{wen2015_wo2}. Organic polymers are transparent, affordable, and flexible; however, they have vibrational absorption peaks beyond the IR band, conflicting with electrochromic device requirements. Key attributes for efficient electrochromic devices include coloration efficiency, easy color management, and rapid responsiveness. One of the most important parameters of electrochromic material, response time, is mostly influenced by the material's absorption and electron exchange capacity\,\cite{GRANQVIST1994_absorption_behind_electrochromism}. Recent studies highlight the advancements in electrochromic device design utilizing conjugated organic polymers and plasmonic nanopixels, which are highly energy-efficient\,\cite{Lv_2017_polymer_redox, peng_2019_nanopixel}. Significant progress has been made in applications like smart windows and eyewear. Recent research on electrochromic devices highlights eco-friendly device implementation and integration into smart displays and wearables, including military camouflage and spectroscopic uses. Their popularity is due to energy-saving features and appealing design.

Challenges like modulation delay and coloration inefficiency limit the vast implementation of commercial electrochromic devices (ECDs). Gaupp \textit{et al.} developed a method to quantify the coloration efficiency of organic electrochromic polymers compared to inorganic ones\,\cite{gaupp_2002_coloration_efficiency}. Coloration inefficiencies originate from charge and ion transport kinetics\,\cite{Ma2017_chemistry_behind_ec}. In any coloration devives, long-term sustainability is crucial, yet many electrochromic systems suffer from low lifetime. Factors influencing ECD durability include electrolyte composition and electrode design. Frequent redox reactions can degrade color-switching properties\,\cite{yang_2020_advances_in_elect}. Additionally, electrolytes can leak or evaporate, risking short circuits or device failure\,\cite{Au_2022_solid_electrolyte}. Moreover, structural integrity can also be compromised by mechanical or chemical stress. Though electrochromic materials-based nanopixels consume low energy, some major issues obstruct the vast commercialization of these devices. A major portion of proposed electrochromic devices exhibit inadequate reversibility. Frequent redox cycles give rise to a phenomenon like electromagnetic hysteresis, referred to as electrochromic hysteresis. This phenomenon is characterized by a delay observed during repeated reverse redox cycles\,\cite{sone2006hysteresis}. Moreover, electrochromic display devices possess a notable characteristic known as the memory effect, wherein individual pixels retain their previous state even in the absence of an external power supply. Hence, undesirable outcomes are observed in frequent refreshing devices, such as televisions or any type of display screen. Hysteresis and memory effects can be attributed to such phenomena. The factors were relatable to the rate of chemical reactions, the mobility of ions, and the capacity of the electrodes to store and discharge electrical charge. Several approaches, such as optimizing the time interval for EC material deposition and modifying the morphology, have been demonstrated to help enhance ion mobility and, consequently, improve device performance\,\cite{LEFTHERIOTIS2008_ion_mobility_enhancement,WU2022_memory_effect}. Recently, plasmonic resonance by metal or metal-like nanoparticles has attracted much attention from researchers and industrial developers. Active plasmonic color generation from light-matter interaction at the nanoparticle edge opens a wide range of scopes for optical device development. The demand for effective plasmon tuning systems has motivated multiple research teams to investigate alternate electrochromic materials, encompassing a diverse range of organic and inorganic materials. In addition, novel nanoparticle configurations are being suggested to enhance plasmonic device performance\,\cite{Akhtary:23_dimer, Nowshin23RINP}. The performance of electrochromic devices can be enhanced by increasing their surface area and improving their ion transport property. Higher conductivity and looser molecular packing of the nanoparticles can effectively contribute to expediting electrochemical reactions\,\cite{Ma2017_chemistry_behind_ec}. 


The permittivity profile of electrochromic material varies with the external electric field applied due to the Kerr effect. Most of the works till now are based on experimental data from ellipsometric measurements. However, there is a scope for further study of the atomistic calculation of EC materials' permittivity for various oxidization states.  Despite recent progress on fast switchable, vibrant organic electrochromic nanopixel design, several perspectives related to color tuning with spherical as well as non-spherical asymmetric shapes with different organic materials have rarely been studied. Moreover, there is a need to extend the color producibility spectrum range by utilizing electrochromic nanoparticles for mirror structure optimization and shell material analysis. The main challenges of existing display technologies that inspired us to study electrochromic core-shell nanopixel display are (i) lack of flexibility, (ii) lack of long-term reproducibility (iii) lower coloration efficiency, (iv) limited lifetime, (v) complex fabrication steps involving toxic elements, and (vi) expensive production process.\\

In this paper, we performed an extensive simulation of NPoM pixel and studied the scopes to enhance the performance and explored various aspects of such pixels. We investigated the color tunability of four electrochromic organic materials and their applicability in electrochromic nanoparticles on the mirror (eNPoM) nanopixel devices. Not only focusing on materials, we emphasized novel nanoparticle design and mirror composition to explore further aspects of the proposed device. In our proposed design, light was tightly confined in the small gap between the nanoparticle and the mirror. Independent adjustment of each nanoparticle was required to function those as an active nanopixel. Such adjustment can be possible by changing the redox state of the polymer shell, which enabled a rapid shift in the resonant scattering color of eNPoM over a wide wavelength range. The main aim of the work was to explore novel designs in organic electrochromic material-based nanopixels. To ease the rapid color-changing cycle in electrochromic materials, we emphasized external stimuli rather than physical parameter alteration. The proposed study focused on the modulation of electrochromic organic conductive polymers via redox reaction. Externally stimulated color-scattering mechanisms rather than permanent particle deformation will open a new era in organic display technology. We determined the scattering cross-section profile for all the electrochromic polymers and represented their color coordinate shift. Moreover, a comparative analysis was conducted on various nanoparticle shapes and on the differences between organic and inorganic electrochromic materials to facilitate material selection. The reproducibility of color generation after subsequent rapid redox cycles of the electrochromic materials will open newer aspects of electrochromic device application diversification.

\section{Methodology}

Permittivity profile of different states resulting from the change in applied voltage of electrochromic materials was investigated in the framework of DFT based first-principles calculations. DFT is a quantum-mechanical (QM) method used to calculate electronic, optical, thermal, and other properties using atomistic simulation which is based on the Khon-Sham equation \cite{Kohn_1964_DFT}, where the energy of a system is defined by a density function independent of external potential.
We comprehensively explored four different organic electrochromic materials- (i)  polyaniline (PANI), (ii) poly 3,4-ethylenedioxythiophene (PEDOT), (iii) polypyrrole (PPy), and (iv) polythiophene (PTh).  

Initially, the unit cells of each electrochromic polymer were created for different oxidization states. For proper modeling of different atomistic structures of different redox states, a structural formula with an atom arrangement was extracted from previously reported experimental works\,\cite{SANCHEZ_2019_book_PEDOT, Ohira2017_PEDOT_Synthesis}. The generated models of unit cells are shown in \textcolor{blue}{Supplementary Material}. We created a polymer chain using the successive repetition of unit cells. For the reduced-state polymer, the unit cell of the reduced state was repeated. Similarly, the oxidized polymer was composed of an oxidized repeat unit. In the case of semi-oxidized state polymers, we considered the reduced state repeat unit for half of the polymer chain, and the rest consisted of the oxidized repeat unit. The reduced states of the built polymer models are shown in Fig.\,\ref{structure} (a). The polymers' semi-oxidized and oxidized states are shown in Supplementary Material.

To model realistic atomic orientation, it is important to predict the three-dimensional arrangement of atoms, which results in a minimum energy level. Hence, we performed geometry optimization before determining the optical and structural properties. The optimization minimizes the net cell volume deviation in consecutive iterative solves. PANI, PEDOT, PPy, and PTh were geometrically optimized until an appropriate convergence threshold was reached. We used CASTEP\,\cite{castep} for geometry calculation. The atomic coordinates and lattice parameters were optimized to minimize energy and force. The convergence threshold for energy deviation was  2$\times$10$^{-4}$ eV/atom. The maximum force on each atom was set at 0.05 eV/\AA. We considered 0.1 GPa the maximum stress on the electrochromic compounds. The displacement after each iteration was limited to 0.002 \AA.

For proper energy calculation, we included both lattice parameters and atomic position optimization in the geometry optimization process using the Broyden-Fletcher–Goldfarb–
Shanno (BFGS) algorithm. To approximate the electron-ion and electron-electron interaction in different states of conductive polymers, Perdew-Burke-Ernzerhof (PBE) approximation of generalized gradient approximation (GGA) functional was utilized in CASTEP\,\cite{castep} package.
DFT calculation needs a dispersion-correlated correction mechanism (DFT-D) to consider the van der Waals force. We used the Grimme method\,\cite{Grimme_2010} for DFT-D. To terminate the self-consistent field (SCF) calculation after a certain convergence, the SCF tolerance was chosen as $2\times10^{-6}$ eV/atom. As the atoms go through a redox cycle, it is obvious that the electron cloud on the valance orbital was not fixed. In each oxidation cycle, the probability of an electron in the valance band significantly decreases. Hence, we did not fix the orbital occupancy in geometry optimization and energy calculation.  Monkhorst-Pack k-grid (2$\times$3$\times$5) was used for PANI, PPy, and PTh to ensure a homogeneously distributed k-point grid in the reciprocal lattice. For PEDOT, a k-grid of (1$\times$4$\times$7) was considered. We added norm-conserving pseudopotential functions to represent the assumed core-packed ions. The cell volumes of these four different materials are tabulated in Table~\ref{table:cell_volume}.

\begin{table}
\small
  \caption{\ Cell volume of different conductive polymers.}
  \label{table:cell_volume}
 \begin{tabular*}{20pc}{@{\extracolsep{\fill}}ll@{}}
    \hline
    Material & Cell Volume ({\AA$^{3}$})\\
    \hline
    PANI & 1770.934145\\
    PEDOT & 1513.654130\\
    PPy & 634.219098\\
    PTh & 2629.606034\\
    \hline
  \end{tabular*}
\end{table}


\subsection{Optical property calculations}

Although we focused on the visible wavelength range for display applications, we calculated the optical properties for a wide wavelength range covering ultraviolet (UV), visible, and infrared (IR) spectra. The optical properties, including complex refractive index and absorption coefficient, for various electrochromic materials at different states were calculated using DFT for the optimized structures mentioned earlier.  Norm-conserving pseudopotentials with appropriate cutoffs were used to describe the electron wavefunctions, allowing proper numerical convergence with reasonable computation time.

The free-electron model can describe the optical properties of plasmonic materials such as Au and TiN. Plasma can be considered a collection of free-conduction electrons. In our study, there was a potential difference between nanoparticles and mirrors in the redox process; proper permittivity approximation for different states of nanoparticles and mirrors was necessary. Here, Drude-Lorentz equation-based mathematical modeling using previously reported experimental data was used to explain their carrier concentration dependent $\epsilon_1$ and $\epsilon_2$ characteristics. We used the Drude-Lorentz formula with two Lorentzian terms to describe how permittivity depends on carrier concentration precisely\,\cite{Islam:23_Plasmon_Tuning_TiN},

\begin{equation}
\varepsilon(\omega)=\varepsilon_{\infty}\mbox{-}\frac{\omega_{\mbox{p}}^{2}}{\omega^{2}\mbox{+i}\Gamma_{D}\omega}\mbox{-k}_{1}\frac{\omega^{2}_{1}}{\omega^{2}\mbox{-}\omega_{1}^{2}\mbox{-i}\gamma_{1}\omega_{1}}\mbox{-k}_{2}\frac{\omega_{2}^{2}}{\omega^{2}\mbox{-}\omega_{2}^{2}\mbox{-i}\gamma_{2}\omega_{2}}.
\label{drude}
\end{equation}
Here, $\epsilon_{\infty}$ is the background permittivity. We assumed a two-sided split system, where one side has fixed ions and the other has mobile electrons. Moreover, we considered the polarization of the ionic cores by any external field, along with the atomic polarization. The background permittivity, $\epsilon_{\infty}$, depends on frequency as polarization does. We simplified Eq. \ref{drude} by taking $\epsilon_{\infty}$ as frequency-independent. The Drude term has a damping coefficient $\Gamma_D$, and the Lorentzian terms have damping coefficients $\gamma_j$ (j=1,2). The plasma frequencies for Drude and Lorentz parts are $\omega_p$ and $\omega_l$ (l=1,2), respectively, which show their resonant frequencies for different transitions. The Lorentzian expansion term has constants $k_i$ (i=1, 2) that indicate the Lorentz oscillation strength and can be obtained by fitting experimental data. The carrier concentration, $n_b$, determines the plasma frequency, $\omega_p$, as follows: 
\begin{equation}
\label{plasma frequency}
\omega_{p}=\sqrt{\frac{\mbox{4}\pi\mbox{n}_{b}\mbox{e}^{2}}{\mbox{m}^{*}}}.
\end{equation} 
Here, an electron’s effective mass is denoted as m$^*$, and its charge is $e$. $n_b$ represents the bulk carrier concentration. As potential was applied to achieve different redox states of shell electrochromic materials, the core plasmonic materials were at various potentials alongside the shell polymer. Therefore, we modeled the potential profile of the core and mirror plasmonic material by its carrier concentration. Hence, the plasma frequency was modulated for different carrier concentrations\,\cite{Islam:23_Plasmon_Tuning_TiN}.

\subsection{FDTD calculations}
The material characteristics extracted from the DFT-based atomistic solution were utilized in the FDTD solver to reflect the change of optical parameters due to the redox cycle. Device structure design was the initial step of FDTD calculation, where an Au block with a dimension of (630nm $\times$ 440nm was set up as a substrate and a mirror for nanoparticles. A sphere with an outer shell was placed on the Au mirror as a nanoparticle. Alongside the sphere, two other different structures, bow tie and gear, with similar core-shell type structures, were investigated here.  Table\,\ref{table:nanopixel_parameters} shows all the device geometrical parameters used.

\begin{table}
  \caption{Geometric parameters of nanopixels with various shapes}
  \label{table:nanopixel_parameters}
 \begin{tabular*}{21pc}{lll}
    \hline
   Structure & Parameter & Dimension (nm)\\
    \hline
  Mirror & Mirror dimension & 630 $\times$ 440$\times$ 15\\
   \hline
  Sphere & Au nanoparticle diameter & 80 \\
   & Electrochromic shell thickness & 10\\
    \hline
  bow tie & Dimension & 154 $\times$ 200 $\times$ 154\\
   & Au thickness &  100\\
   & Electrochromic shell thickness & 10\\
    \hline
  Gear & Metal radius & 70\\
   & Au inner radius & 30\\
   & Au thickness & 200\\
   & No. of teeth & 20\\
   & Teeth face width & 15\\
   & Electrochromic shell thickness & 10\\
   \hline
  \end{tabular*}
\end{table}

We analyzed the nanostructures studied in this study using FDTD method. In the FDTD method, continuous recursive upgradation of the magnetic and electric fields gave the complete electric as well as magnetic field profile inside the structure in time domain\,\cite{sullivan2013electromagnetic}. 
Chirped z\textendash transform, czt as expressed in Eq. \eqref{czt} was used to get frequency spectrum from the time-domain field signal.
\begin{equation}
\label{czt}
    czt (E,x,k)=\sum_n{E_{x}[n]e^{ix[n]k[m]}}
\end{equation}
Here, ${E_{x}}$ is the electric field magnitude along x direction, ${n}$ represents the refractive index of the medium and $k$ denotes the frequency index. Index characterization of electrochromic materials was performed utilizing our DFT-calculated n and $\kappa$ data. The refractive index of Au was calculated using the Drude-Lorentz formula described earlier. We calculated the scattering and absorption spectra of all four electrochromic materials studied here. As our study focused on scattering calculation and local field enhancement, a total-field scattered-field (TFSF) source was employed to excite the nanoparticle. This source separated the scattered light from the nanoparticle from the light emitted by itself. The nanoparticle and associated structures were kept inside the TFSF region to ensure our calculation's accuracy. We set the source wavelength range in the visible spectrum (400 -- 800nm) as our proposed model was designed for visible light applications. The permittivity of electrochromic material was calculated using DFT, gold's permittivity profile was calculated by solving the Drude -Lorentz equation, and the surrounding medium of the simulation arena was considered filled by air with a permittivity constant = 1. The Au-air interface is set as z = -r, and a nanosphere is positioned at the interface with the center coordinate being (0, 0, 0) in which \textit{r} is the radius of nanospheres.  In plasmonic structures, the interaction between the electromagnetic field and nanoparticle edge produces a strong electric and magnetic field enhancement in the junction of different dielectric materials. Frequency domain field monitors were placed around the simulated structure to capture all sides' scattered light. Afterward, the scattering cross-sections from all six sides were accumulated to determine the net scattering cross-section. This was performed utilizing six independent monitors to form a closed box around the core-shell structure.  These monitors measured the net power flowing into/out from the structure, and the cross-section was determined by normalizing it to the source intensity. To avoid the inclusion of indecent beam energy, all the above-mentioned monitors were placed outside of the TFSF source region. Moreover, extra care was taken to place the monitors to avoid excluding the core-shell structure from the monitor arrangement. As our study included core-shell structure and the three-dimensional shape played a great role in performance, we conducted 3-D simulations for all studied structures. So, our calculated cross-sections were in $m^2$ unit.  On the other hand, the absorption cross-section was calculated using a similar procedure, except the monitors were placed inside the TFSF source region. The TFSF source works best in uniform mesh scheme, we kept 6th order uniform mesh throughout the structure except electrochromic material-mirror junction area. 

\begin{figure}[h]
\centering
\includegraphics[width=0.5\textwidth]{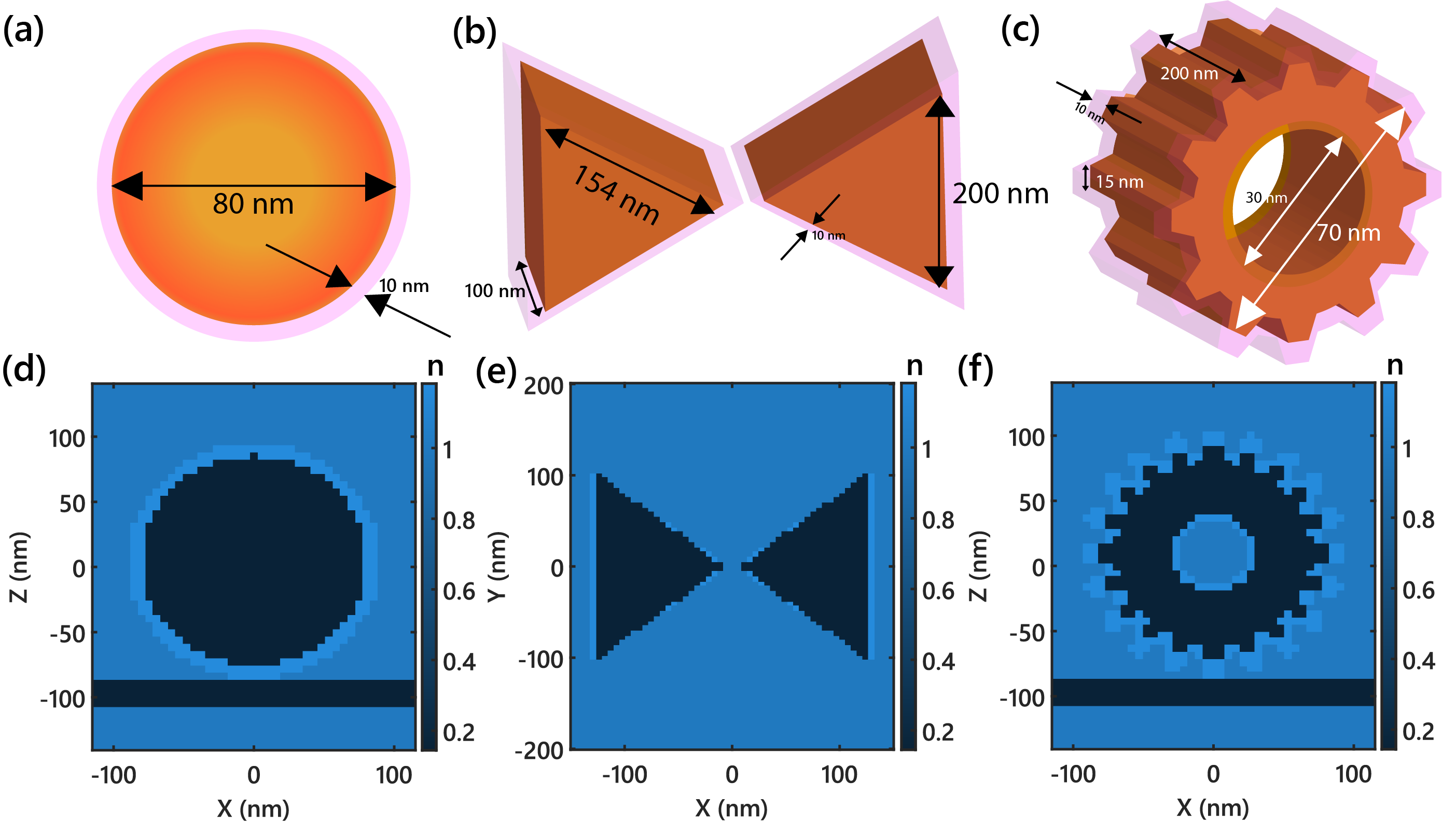}
\caption{3d illustration of Au core- EC shell structures used in the study with geometrical dimensions (a) sphere, (b) bow tie, and (c) gear. Index plot after mesh override for (d) sphere (e) bow tie, and  (f) gear core-shell structure. The illustrated nanoparticles presented here are not drawn to the scale.}
\label{index}
\end{figure}

In the electrochromic-mirror junction, it is essential to maintain a sufficiently dense mesh grid for proper index allocation and calculation. Otherwise, comparatively coarser mesh in metal edges results in inaccurate results as there is significant permittivity change at that region. Hence, we added a special meshing grid with a 5nm resolution in all three dimensions in that area. To avoid divergence and other complexities in FDTD calculation, a gap was maintained between the TFSF source and monitors. Moreover, the simulation region was chosen sufficiently large so that most scattered light could pass through the monitors and be included in the calculation. Another important aspect is the selection of a proper mesh override index. All of our structures contain a core-shell pattern. It is cumbersome to model such core-shell structure, especially in complex structures like gear. We chose a proper mesh override index to simulate the inner metal nanoparticle with an outer electrochromic shell surface. Moreover, the mesh override regions were covered by the metallic mirror. Mesh override allowed separate index modeling for the core and shell structures. Fig.\,\ref{index} shows the index profile after mesh overriding. After incorporating all the calculated material data into the NPoM structure, the structures' scattering profile, absorption, and other necessary profiles were studied. After successfully implementing a tunable electrochromic device, approaches for further enhancement of the device structure, electrode, and charge carrier layer were investigated, and necessary optimization was performed to make an efficient, broad-range color reproducible pixel. In this device, shell geometry optimization is vital for better scattering. So, we optimized the shell geometry and determined the scattering maxima for each structure studied here, such as sphere, bow tie, and gear.
The core and shell are symmetrical in spherical nanoparticles, and symmetric boundary conditions were applied on the symmetric axes. On the other hand, bowtie and gear-shaped electrochromic nanopixels were simulated using perfectly matched layer (PML) boundary conditions on asymmetric axes. PML layers have some issues. The most prominent one is the possible inclusion of reflected light from the layers. To avoid this incident, we deployed eight separate layers in stretched coordinate PML to avoid this incident. Afterward, we performed the scalability of the proposed devices by simulating an array of nanopixels.

\subsection{Chromaticity calculations}

After the DFT and FDTD calculations, we analyzed the color-switching capability of the proposed device by estimating the CIE 1931 XYZ chromaticity diagram. The color space specified by the International Commission on Illumination (CIE) in 1931 for color matching functions where a color element is shown graphically in a CIE 1931 chromaticity diagram. The horseshoe-shaped plot indicates the range of colors produced by pure monochromatic lights. It also describes the correlation between light wavelengths and perceived colors. A diagonal line known as the white point line, which goes through the point (1/3, 1/3), which corresponds to the equal-energy white, divides the diagram into two sections. The positive region is the area above the line, and the negative region is below the line. A color that can be represented by varying the ratios of three primary colors red, green and blue (RGB) can be represented by any point inside the diagram. The point's coordinates, represented by x and y, are the chromaticity coordinates. The following formulas relate the chromaticity coordinates to the tristimulus values X, Y, and Z:

\begin{equation}
    \mbox{x=}\frac{\mbox{X}}{\mbox{X+Y+Z}}
    \label{CIE_1931_X_cordinate}
\end{equation}

\begin{equation}
    \mbox{y=}\frac{\mbox{Y}}{\mbox{X+Y+Z}}
    \label{CIE_1931_Y_cordinate}
\end{equation}

Several color-representing standards, such as RGB, CMYK, and Lab*, can be used in chromaticity diagrams. Further analysis of the CIE 1931 chromaticity diagram can also measure the color temperature, color rendering index, and color difference of light sources or colors.

We used a MATLAB script to plot our FDTD calculated data in the CIE 1931 chromaticity diagram. We considered the D65 illuminant which mimics the spectral power distribution of natural sunlight, as the light source in plotting the CIE 1931 diagram. It was employed in various simulations, color optimizers, plastic production, textiles, inks, and the automobile industry. It is very useful in color-matching applications. Its corresponding color temperature is around 6500 K, indicating a bluish tint of illumination. Our main goal was to determine the coordinates of electrochromic nanopixels in CIE 1931 color space. Utilizing equations \ref{CIE_1931_X_cordinate} and \ref{CIE_1931_Y_cordinate}, we determined the xy coordinate from the RGB values. These RGB values gave a specific color for a given state of electrochromic nanopixel, shown in the color palette alongside the CIE 1931 chromaticity diagram. In the chromaticity diagram, the color coordinates of all three redox states of electrochromic material are given to give an idea of the color shifting range in the color space. 

\section{Results and discussion}
We focused on redox cyclic refractive index variation using conductive organic polymers. It is a comparatively more feasible technique to implement organic, conductive polymer-based electrochromic devices than inorganic ones. Moreover, this approach resulted in easy-to-tune resonance switching. Initially, we simulated our proposed nanopixel over Au mirror to investigate the redox switching capability of organic electrochromic polymers studied. Afterwards, for more detailed analysis and comparative study, we optimized our nanopixel geometry and added a 15nm TiN layer over Au substrate.

\begin{figure*}
\centering
\includegraphics[width=1\textwidth]{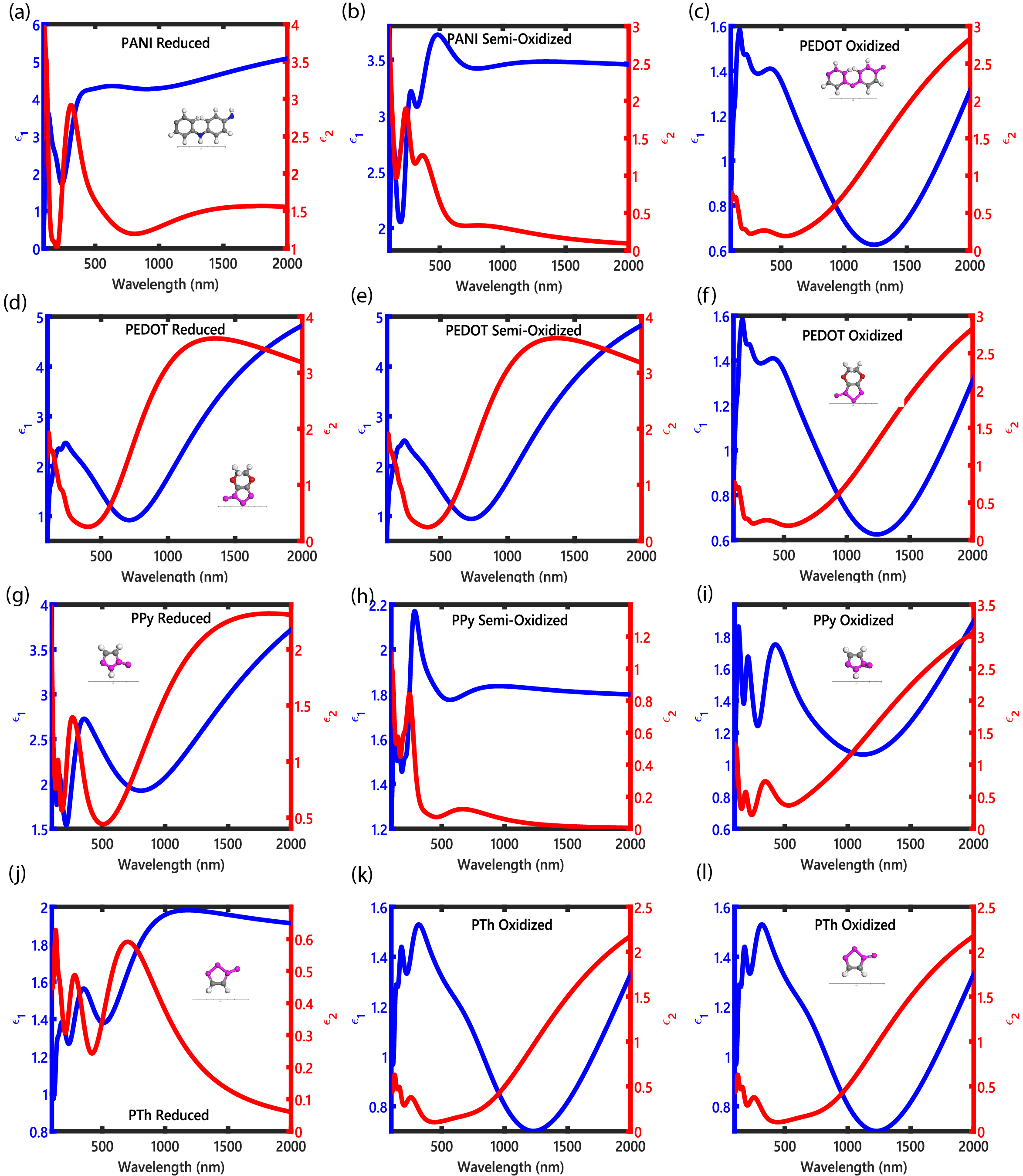}
\caption{Calculated $\epsilon_1$,$\epsilon_1$ spectra for (a) PANI-reduced, (b) PANI-semi oxidized, (c) PANI oxidized, (d) PEDOT-reduced, (e) PEDOT-semi oxidized, (f) PEDOT oxidized, (g) PPy-reduced, (h) PPy-semi-oxidized (i) PPy-oxidized, (j) PTh reduced, (k) semi-oxidized Pth, and (l) oxidized PTh using DFT.}
\label{dielectric_function}
\end{figure*}

\subsection{Redox cycle of EC materials}
PANI, PEDOT, PPY, and PTh have gained much attention in flexible device fabrication. Among them, PEDOT can be considered as a derivative of EDOT, more explicitly Hydroxy methyl-EDOT. All of these polymers are easy to synthesize and commercially available. Like other conductive polymers, PANI, PEDOT, PPY, and PTh have conjugated backbones with $\sigma$ bonds, and overlapping $\pi$ bonds are shown in Fig. S1 of Supplementary Material. The polymer building $\pi$ bonds are loosely attached to the atoms, resulting in the delocalization of electrons. Therefore, electrons can easily move around the polymer chain, which is the main reason behind their conductivity. Utilizing this property, we explored the change of optical property, local field enhancement, and, ultimately, the color switching on the proposed nanopixel through a redox cycle of the mentioned polymers. Among the studied 4 polymers, PANI is the most widely used organic electrochromic material so far. The most widely utilized mechanism for synthesizing PANI and similar organic conductive polymers is oxidant polymerization, in which aniline reacts with an acidic material as a neutralizing agent and gets polymerized when one drop of an oxidizing agent, such as ammonium persulfate (NH$_4$SO$_4$) is added at different temperatures\,\cite{pani_redox_mech}. The oxidation and reduction process of PANI is continuous. Nonetheless, three major states of PANI are (i) leucoemeraldine (C$_{6}$H$_{4}$NH) -- colorless and white/clear, (ii) emeraldine ([C$_6$H$_4$NH]$_{2}$[C$_{6}$H$_{4}$ N]$_{2})_{n}$) -- blue as emeraldine base and green as emeraldine salt, and (iii) pernigraniline (C$_{6}$H$_{4}$N)$_{n}$) -- Blue/violet. Among them, the emeraldine form of polyaniline, often called emeraldine base (EB), is neutral (se Supplementary Material). If doped (protonated), it is called emeraldine salt (ES), where the imine nitrogens get protonated by an acid. This protonation process results in reversible oxidation-reduction\,\cite{PANI_Protonation_regina_2019} which ultimately leads to the chromaticity change in polymers.

Poly 3,4-ethylenedioxythiophene (PEDOT), or ethylenedioxythiophene (EDOT), is a 3,4-di-substituted thiophene. PEDOT is a conducting polymer with numerous uses in both biomedicine and industries\,\cite{Rahimzadeh2020_pedot_introduction}. This polymer has numerous advantageous properties compared to other conducting electrochromic materials. The oxidative polymerization mechanism of PEDOT can be divided into three steps. (i) \textit{Oxidation of EDOT monomers:}
Oxidizing agents such as ammonium per sulfate, oxidizes the 3,4 ethylenedioxythiophene (EDOT) monomers to produce radical cations\,\cite{Volk_2021_PEDOT_oxidation_processs}. (ii) \textit{Dimerization of radical cations:}
Bi-thiophene units are created when the radical cations become dimerized. Afterward, these dimers can combine with more radical cations to create oligomers and, ultimately, polymers (PEDOT)\,\cite{Volk_2021_PEDOT_oxidation_processs}. (iii) \textit{Deprotonization of polymers:}
Strong bases, such as sodium hydroxide deprotonate the PEDOT polymers to make the neutral PEDOT chains. PEDOT's electrical and optical characteristics are basically governed by its conjugated backbone, which alternates between single and double bonds in different states\,\cite{Volk_2021_PEDOT_oxidation_processs}.

Polypyrrole (PPy) is an intriguing organic polymer. Pyrrole undergoes oxidative polymerization to yield Polypyrrole (PPy). PPy is an intrinsically conducting polymer, setting it apart from traditional organic polymers that are typically insulating. Its ability to conduct electricity makes it a valuable material in various fields. PPy films are yellow in their reduced state but darken due to oxidation. Our simulation result reflects this property, as shown in Fig.\,\ref{TiN_color_Switching}. PPy is often described as a “quasi-unidimensional polymer” because of some crosslinking and chain hopping inside its backbone.

The following chemical equation can represent the redox reaction of Polypyrrole, PPy, 
\begin{equation}
\mbox{PPy(red)+xA}^-\longleftrightarrow\mbox{PPyxA(ox)+xe}^-,    
\end{equation}
where $A$ is the dopant ion, such as Cl, SO$_4$, ClO$_4$, or NO$_3$, and $x$ is the degree of oxidation. The reduced state of PPy contains $\sigma$ bonds between monomer units, which allow the polymer to freely rotate, which results in different conformational structures due to external effects. The oxidation state of PPy can vary from reduced to fully oxidized, depending on the applied potential or the concentration of the oxidizing agent. In our study, we focused on the earlier one. PPy has a bandgap of around 3.16 eV\,\cite{Pang_2020_PPy_band_gap}. The oxidation process increases its conductivity and decreases the band gap, making it more conductive and less transparent. This oxidation process changes its color from yellow to blue or black, depending on the degree of oxidation and the excitation applied. Interestingly, the redox reaction of PPy is reversible. Hence, PPy can be switched back and forth between its oxidized and reduced forms by changing the polarity of the electric field or the oxidant concentration of the solution.

Polythiophene is the polymerized form of thiophene containing sulfur heterocycle\,\cite{TOURILLON_1982_Pth_introduction}. This conductive polymer can be synthesized by chemical oxidative polymerization\,\cite{Chetia2023_Pth_synthesis}. Polythiophene becomes oxidized by electron acceptors, and then the polymer becomes more conductive. The conductivity can be raised up to 1000 S/cm\,\cite{McCullough_1993_PTh_conductivity}. Bipolarons produced during the oxidation process are responsible for its increased conductivity. Two thiophene units together form the bipolaron unit, which can travel along the polymer chain. In our study, the polymer was oxidized and reduced frequently, so the bipolarons are created and destroyed continuously. When the electron donor gets prominent (in negative voltage), the bipolarons go back to their neutral state and complete the redox cycle.

\subsection{Optical properties  of different redox states}
The refractive index and extinction coefficient are two fundamental optical properties that dictate the performance of polymers in numerous applications. Understanding the interplay between these parameters across the wavelength spectrum is vital for fine-tuning polymer optical properties to specified requirements. The refractive indices for the different oxidation states of the materials studied were calculated to understand their optical property transitions through the oxidization process. Fig. \ref{dielectric_function}  presents the DFT calculated dielectric function of the studied electrochromic polymers: PANI, PEDOT, PPy, and PTh in different oxidation states (Reduced, Semi-Oxidized, and Oxidized). In all sub-figures, the blue curve represents the real part of permittivity, indicating the polymer's charge-storing ability and the red curves present the imaginary part of permittivity, which is directly related to material absorption. Due to the change in atomistic configuration, the oxidation and reduction process significantly affects both $\epsilon_{1}$ and $\epsilon_{2}$.  Both for PANI and PPy, there are strong peaks of  $\epsilon_{2}$ at lower wavelength regions, which results in absorption peaks at lower wavelength regimes shown in Fig.\ref{acs}. This can be attributed to charge carrier confinement and possible localized electron states.  In oxidized states, broad absorption can be expected near IR range, which is beyond the scope of this study. Fig. S4 of  Supplementary Material depicts our key findings from DFT calculation, which indicate that for PANI, both n and k decrease with oxidation, contrasting with the observed increase in these values for PEDOT, PPy, and PTh. These results can be explained in the context of conjugation length and backbone planarity, which influence the delocalization of $\pi$-electrons and, ultimately, the optical properties. Conjugation length depends on the number of repeating units in a polymer chain that participate in the delocalization of $\pi$-electrons. It has a prominent impact on the polymer's electrical conductivity, optical characteristics, and electronic properties. If the conjugation length increases, the polymer's conductivity increases, and its band gap decreases. However, structural determinants, including kinks, twists, and disorder (including defects) in the polymer chain, limit the conjugation length\,\cite{Kürti_1987_conjugation_length}. A polymer's conjugation length can be determined using a variety of techniques, including transport models, quantum chemical calculations, UV-vis spectroscopy, and Raman spectroscopy\,\cite{Namsheer_2020_rsc_advence_conducting_polymer_review}. In our study, we chose conducting polymers with long conjunction lengths. This resulted in a significant deviation in permittivity, as shown in Fig.\,\ref{dielectric_function} (a)-(l). The atomistic models for reduced and oxidized states used in DFT calculations are shown in the inset. 

In our study, the nanopixel scheme was focused on the rapid shift of resonant scattering color utilizing the switching of the charge state of the entire polymer shell. The refractive index (n) values in lower wavelength regions were higher than the extinction coefficient ($\kappa$), signifying higher light bending than absorption. As the wavelength increased, $\kappa$ values exhibited a gradual ascent, converging with and surpassing n values. This crossover suggests a transition from dominant refractive behavior to significant absorptive behavior, a critical consideration for designing devices intended to operate across a wide spectral range.

\subsection{Scattering cross-sections}

\begin{figure}
\centering
\includegraphics[width=0.5\textwidth]{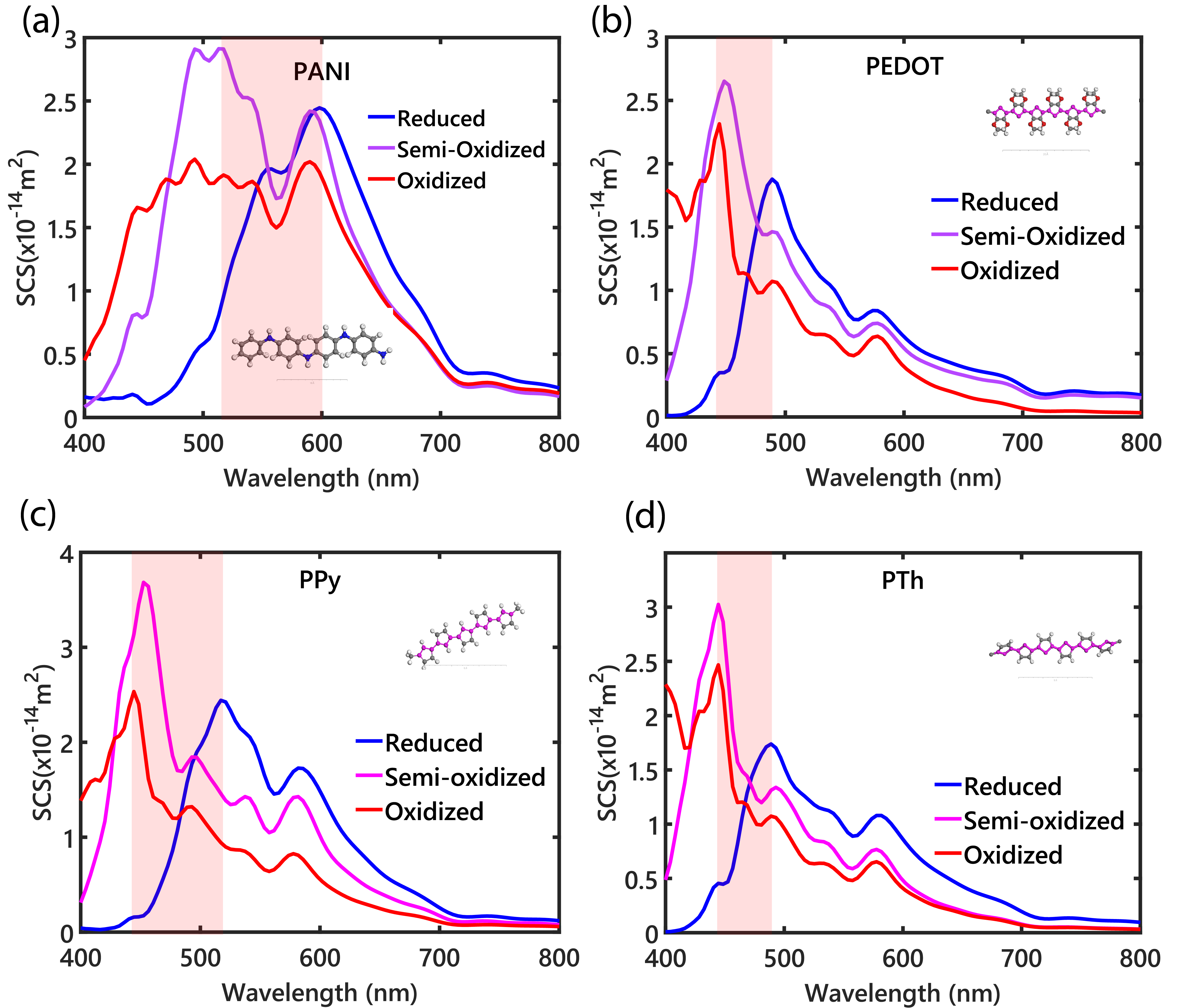}
\caption{Scattering cross-section at different oxidation states of (a) PANI, (b) PEDOT, (c) PPy, and  (d) PTh.}
\label{scs_plot}
\end{figure}

\begin{figure}
\centering
\includegraphics[width=0.5\textwidth]{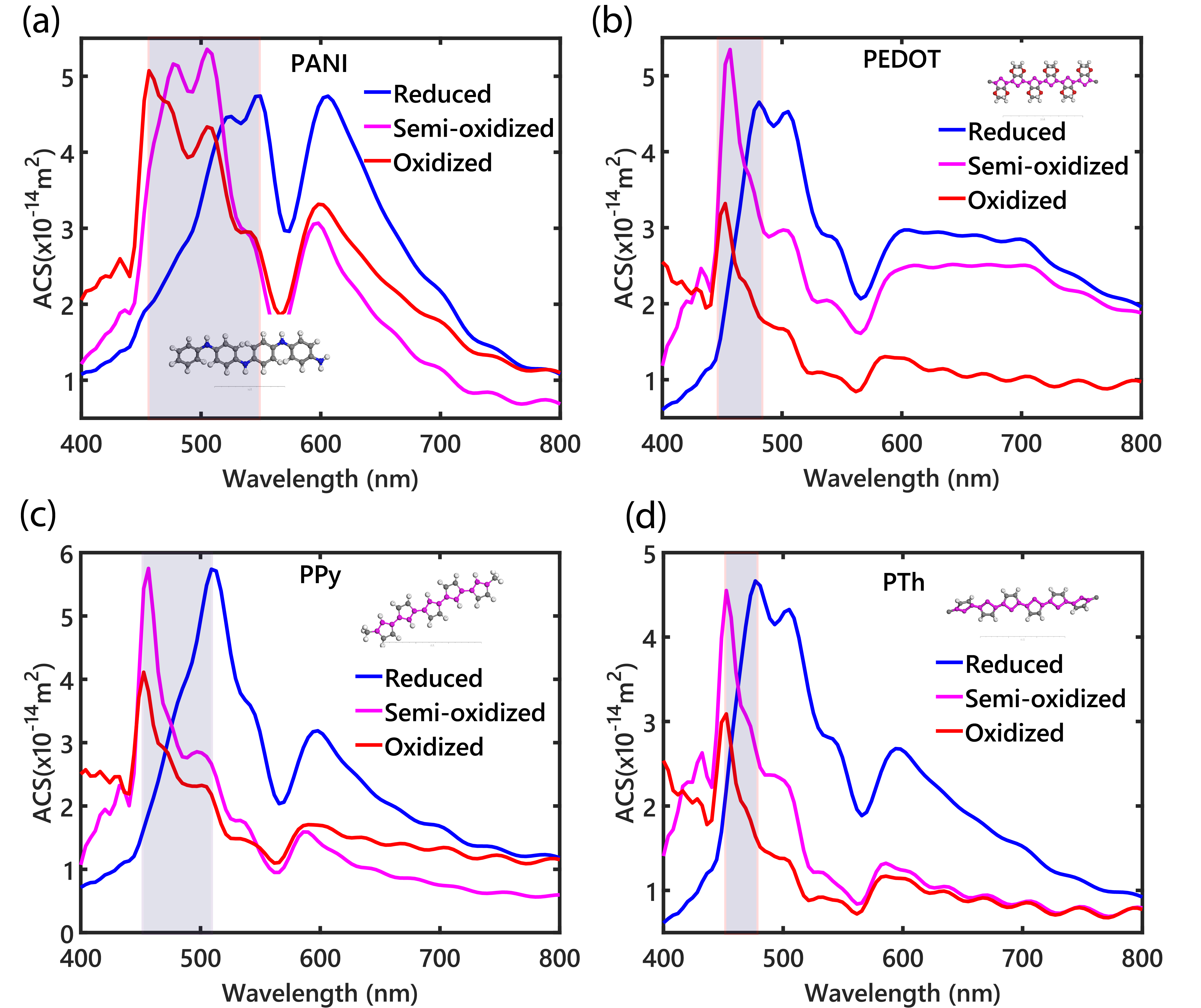}
\caption{Absorption cross-section at different oxidation states of (a) PANI, (b) PEDOT, (c) PPy, and  (d) PTh.}
\label{acs_plot}
\end{figure}

By changing the surrounding refractive index of the nanoparticle, the local surface plasmon resonance (LSPR) peak can be shifted\,\cite{Shao_2018_plasmon_shift, Islam:23_Plasmon_Tuning_TiN}. This shift leads to a change in the scattering profile. Scattering cross-section refers to the total scattered power divided by the incident power per unit area. It is a common practice to normalize the cross-section measurement to the particle size, especially when the particle size is swept over for optimization. However, we didn't normalize the cross-section parameter to particle shape to compare various shaped nanoparticles' scattering tunability in the same scale. The change in resonant wavelength, $\delta\lambda$ is directly linked to dielectric permittivity slope $d\epsilon_{m}/d\lambda$ as below\,\cite{peng_2019_nanopixel,peng_2022_electrochromic},

\begin{equation}
    \frac{\Delta\lambda}{\Delta\mbox{n}}=\frac{\mbox-{2}\chi\mbox{n}}{(\mbox{d}\epsilon_{m}\mbox{/d}\lambda)_\lambda}.
    \label{peak shift}
\end{equation}

Here, n and $\chi$ represent the refractive index of the surrounding medium and the shape factor of the nanoparticle, respectively. To obtain maximum color shifting, maximizing the $\delta \lambda$ value is required to make a tunable nanopixel over the visible spectrum. The external active electrochromic medium paves the way to achieve this. By applying external stimuli on an electrochromic shell, we can tune the surrounding dielectric permittivity around the nanoparticle, which results in resonance shift. In our study, we have calculated the scattering cross-section (SCS) of all the stated electrochromic materials. The SCS profiles of all four electrochromic materials are shown in Fig.\,\ref{scs_plot}. In the case of PANI and PEDOT, the SCS peak blue shifts in the oxidation process. However, it shows a reverse effect in the case of PPy. Among the electrochromic materials, PEDOT gives maximum shift in scattering, more than 50nm shift in the redox cycle, shown in Fig.\,\ref{scs_plot} (b). The wavelength and the magnitude of the scattering cross-section also significantly changed. Peak amplitude of scattering cross section was found dependent on oxidation state. In all the four studied polymers, we got the highest peak of scattering cross section in semi-oxidized state. The cross-section area of scattering was found to be significantly variable on the oxidization state, which is tabulated in Table \,\ref{scs_data}.   In reduced state, conductive polymers have a more uniformly ordered conjugated $\pi$ bond chain which is suitable for better absorption. However, when it got partially oxidized, there was a mix of reduced polymer chain and oxidized polymer chain, resulting in sub-domain variation of permittivity along the chain. Therefore, it can be considered as a mixture of doped (oxidized) and intrinsic (reduced) segments of polymer, which is analogous to semiconductor materials. Moreover, in such a situation, the shell roughness increases due to polymer swelling. Due to the mentioned facts, the amount of Mie scattering on the shell surface got increased. We have simulated the oxidized state of both core and shell structures by imposing their different permittivity profile. In the oxidized state, the nanoparticle and electrochromic shell were considered high E-field regions with respect to the mirror structure.  All the SCS profiles calculated here were derived from the simulation results, utilizing the Au nanosphere with an electrochromic shell layer. However, other possible nanoparticle shell structures were also studied, and the findings from those studies are described in the subsequent section.\\
Any display device must have a pixel device capable of producing basic colors: red, green, and blue (RGB). In Fig.\,\ref{scs_plot}, we can see that a significant range of color spectra can be tuned by changing the oxidation state of the electrochromic shell structure. As previously discussed, the oxidized and reduced states were incorporated by imposing an external electric field on the electrochromic shell material with respect to the mirror. Practical implementation of such NPoM structure on ultrathin film has already been demonstrated by Peng \textit{et al.}\cite{peng_2020}.

\begin{table}
    \centering
\caption{ Peak SCS values in different oxidation states of the polymers studied.}
\label{scs_data}

    \begin{tabular}{cccc}
        \hline
        \textbf{Material}&\multicolumn{3}{c}{\textbf{SCS ($\times 10^{-14}$ m$^2$)}} \\ 
        \cline{2-4}
        &\textbf{Reduced} & \textbf{Semi-Oxidized} & \textbf{Oxidized} \\ \hline
        PANI & 2.44& 2.91& 2.03\\
        PEDOT & 1.87& 2.65& 2.31\\ 
        PPy & 2.44& 3.68& 2.53\\ 
        PTh & 1.73& 3.02& 2.48 \\ \hline
    \end{tabular}

\end{table}

\subsection{Local field enhancement}
\begin{figure*}
\centering
\includegraphics[width=1\textwidth]{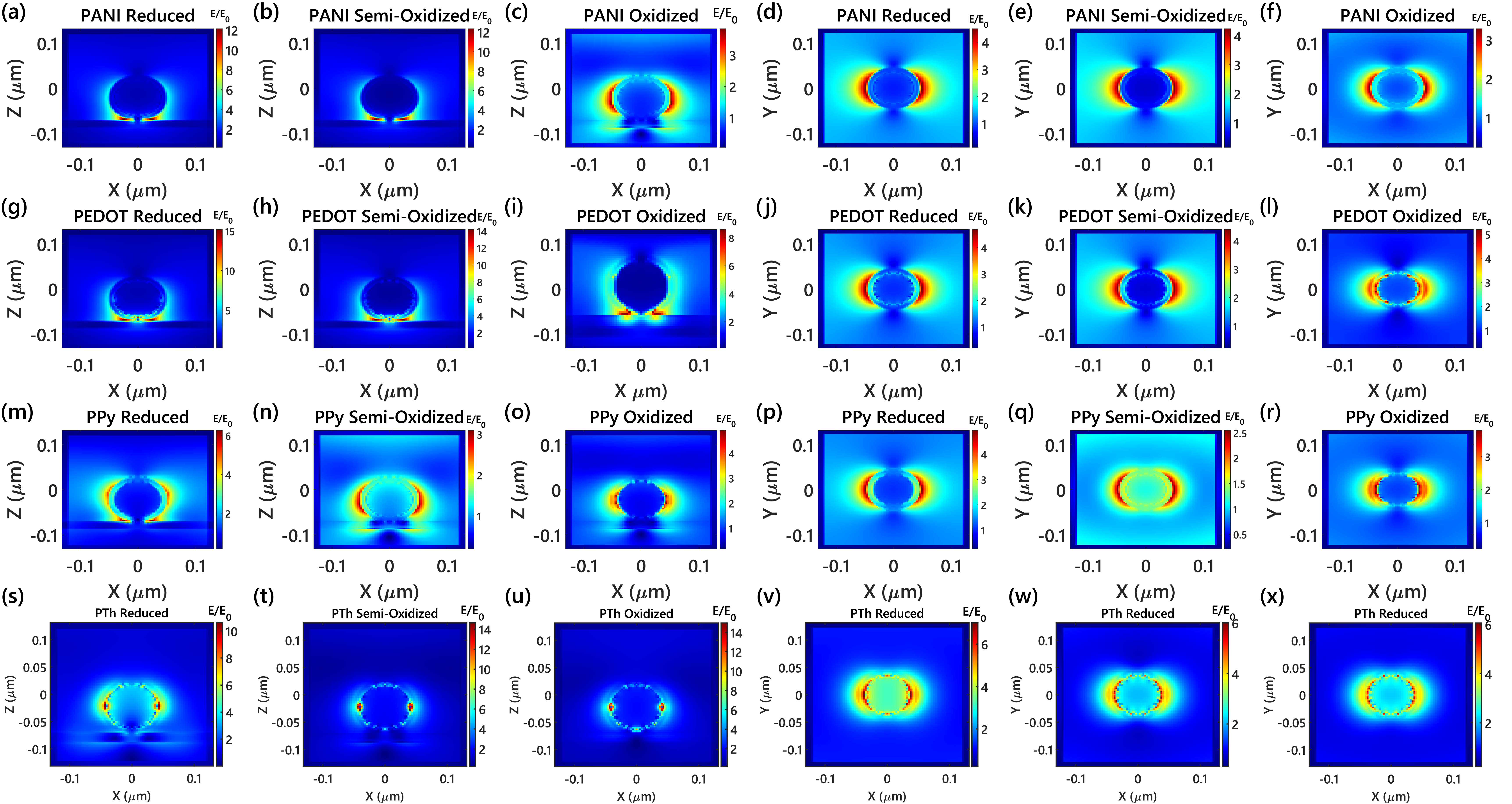}
\caption{Near-field electric field distribution around the electrochromic shell of three different states of oxidization; PANI (a-f), PEDOT (g-l), PPy (m-r), PTh (s-x) y normal view and, (d, e, f, j, k, l, p, q, r, v, w, x): z normal view.}
\label{E-field}
\end{figure*}

Plasmonic structures can confine and enhance the electromagnetic field around them. Electrochromic materials, on the other hand, can change their optical properties in response to external electric field\,\cite{Kim2023_electrochromic}. In our study, these two concepts are integrated into a single nanopixel, where the electrochromic effect can easily modulate the local field enhancement provided by plasmonic nanostructures. Proper evaluation of local field distribution on different electrochromic materials bears much importance for studies on light-matter interaction in nanopixels such as Raman spectroscopy, fluorescence study, and photovoltaic applications\,\cite{Yu2019_field_enhancement}. The nanospheres studied here had a diameter of 80nm with a 20nm electrochromic shell. This introduced a 20nm gap between the Au sphere and the Au mirror. The moderately smaller bandwidth observed in Fig.\,\ref{scs} is evidence of high coupling. This phenomenon ensured a minimum of 3 times electric field enhancement in the local field. The electric field distribution in XZ and XY planes was measured through the center of the Au nanoparticle at the wavelength point, which is equal to the peak scattering wavelength. All the calculated E-field profiles are shown in Fig.\,\ref{E-field}. PANI and PEDOT core-shell on the Au nanosphere efficiently confined the electromagnetic fields in the Au-Electrochromic Layer-Au junction in the sub-wavelength scale. One important notable thing in this figure is the abrupt change of the electric field profile on the borders of the electric field profile. This is because the TFSF source area has both source illumination as well as scattered fields; on the other hand, the region beyond the TFSF area has only scattered fields. This makes the TFSF source area border have a sudden change of electric field. PEDOT shell layers demonstrated the most significant local field enhancement in our study.

Moreover, its maximum color tunability made it a promising candidate for wide-range color-producing pixel elements. In such applications, the emitting efficiency is also a major determinant of overall performance. Moreover, we can see that the electric field enhancement suddenly dropped from a semi-oxidized state during the full oxidization process. In the case of PEDOT and PANI, it dropped from 14 to 8 (42\%) and 14 to 4 (71\%), respectively. This indicates a possibility of utilizing this property as a meta-state switch. The plasmonic resonant frequency is quite dependent on two prominent factors. One is the dielectric constant, and the other is the surface property of metal particle\,\cite{Yu2019_field_enhancement}. To evaluate the effect of oxidation on the field profile, we have simulated all three oxidation states, keeping the geometry constant. Thus, it can be seen from Fig.\,\ref{E-field} that the intensity of the localized field can be controlled dynamically via the redox cycle. This can lead to new ways of manipulating light at the nanoscale range for high-resolution display devices and non-display applications such as sensing, imaging, and photovoltaics.

\renewcommand{\topfraction}{.85}
\renewcommand{\bottomfraction}{.7}
\renewcommand{\textfraction}{.15}
\renewcommand{\floatpagefraction}{.66}
\renewcommand{\dbltopfraction}{.66}
\renewcommand{\dblfloatpagefraction}{.66}
\setcounter{topnumber}{9}
\setcounter{bottomnumber}{9}
\setcounter{totalnumber}{20}
\setcounter{dbltopnumber}{9}

\begin{figure*}
\centering
\includegraphics[width=1\textwidth]{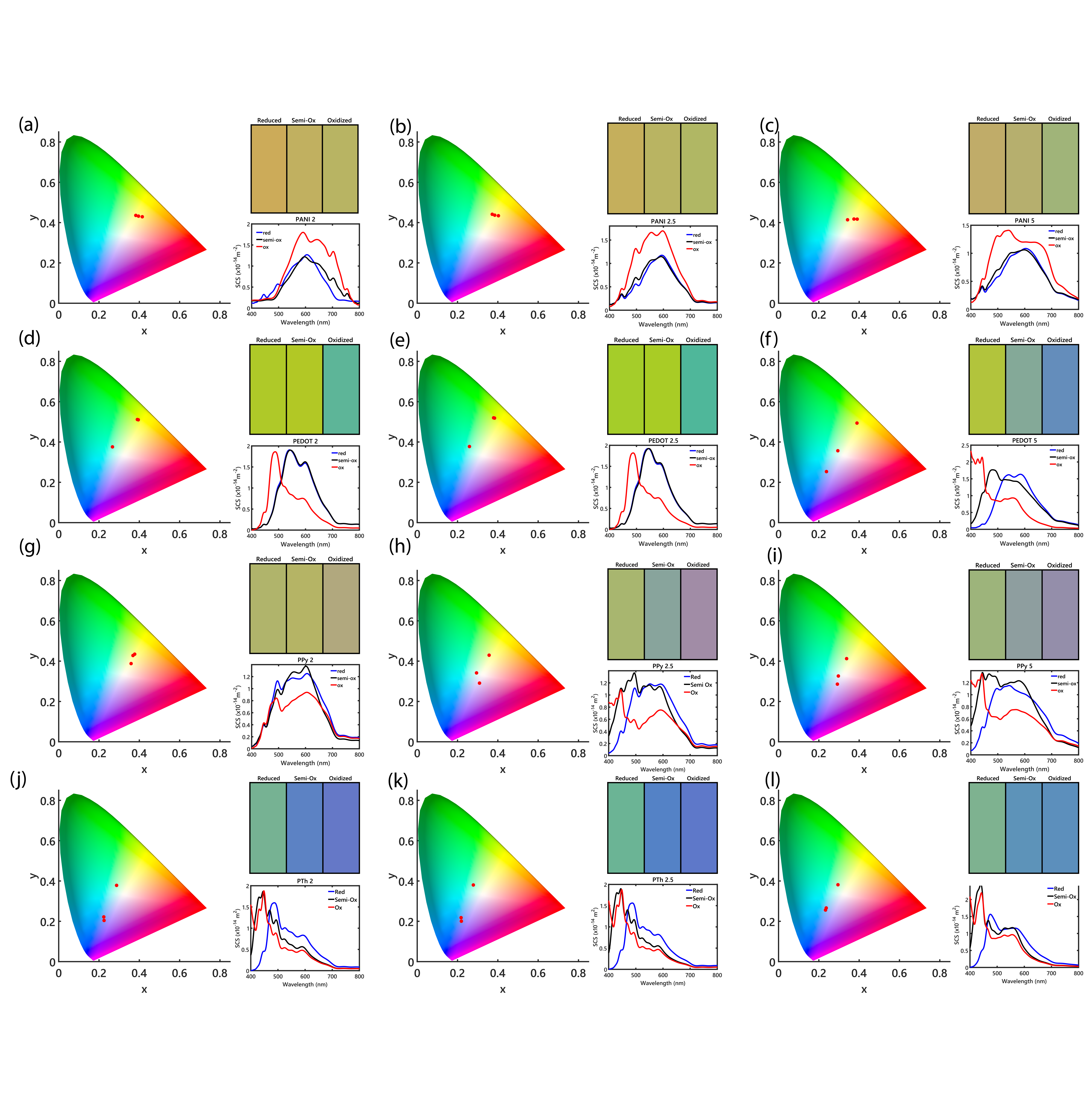}
\caption{Color switching for three different oxidation states of (a-c) polyaniline (PANI), (d-f) 3,4-ethylene dioxythiophene (PEDOT), (g-i) Polypyrrole (PPy), and (j-l) Polythiophene (PTh) shell on Au nanosphere. Different levels of excitations were applied to the TiN upper mirror. The color coordinate shift in CIE 1931 color space, corresponding color pallet, and scattering cross-section is shown for different redox states at different excitations.}
\label{TiN_color_Switching}
\end{figure*}

\begin{figure*}
\centering
\includegraphics[width=1\textwidth]{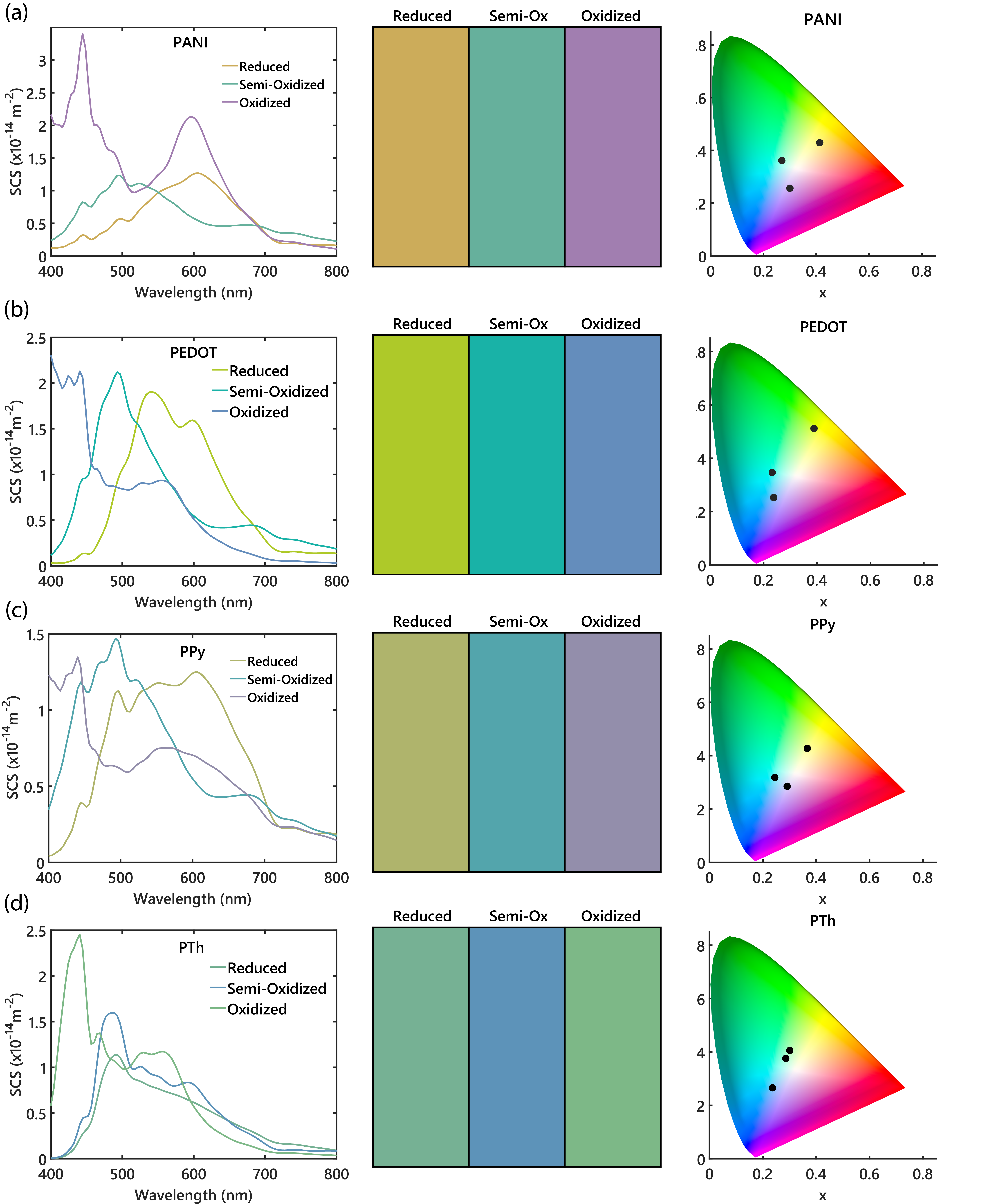}
\caption{Color production via gradient field application on both TiN and Au mirror for electrochromic shell made of (a) PANI, (b) PEDOT, (c) PPy, and (d) PTh.}
\label{TiN_layer_scs_max}
\end{figure*}

\begin{figure}
\centering
\includegraphics[width=0.5\textwidth]{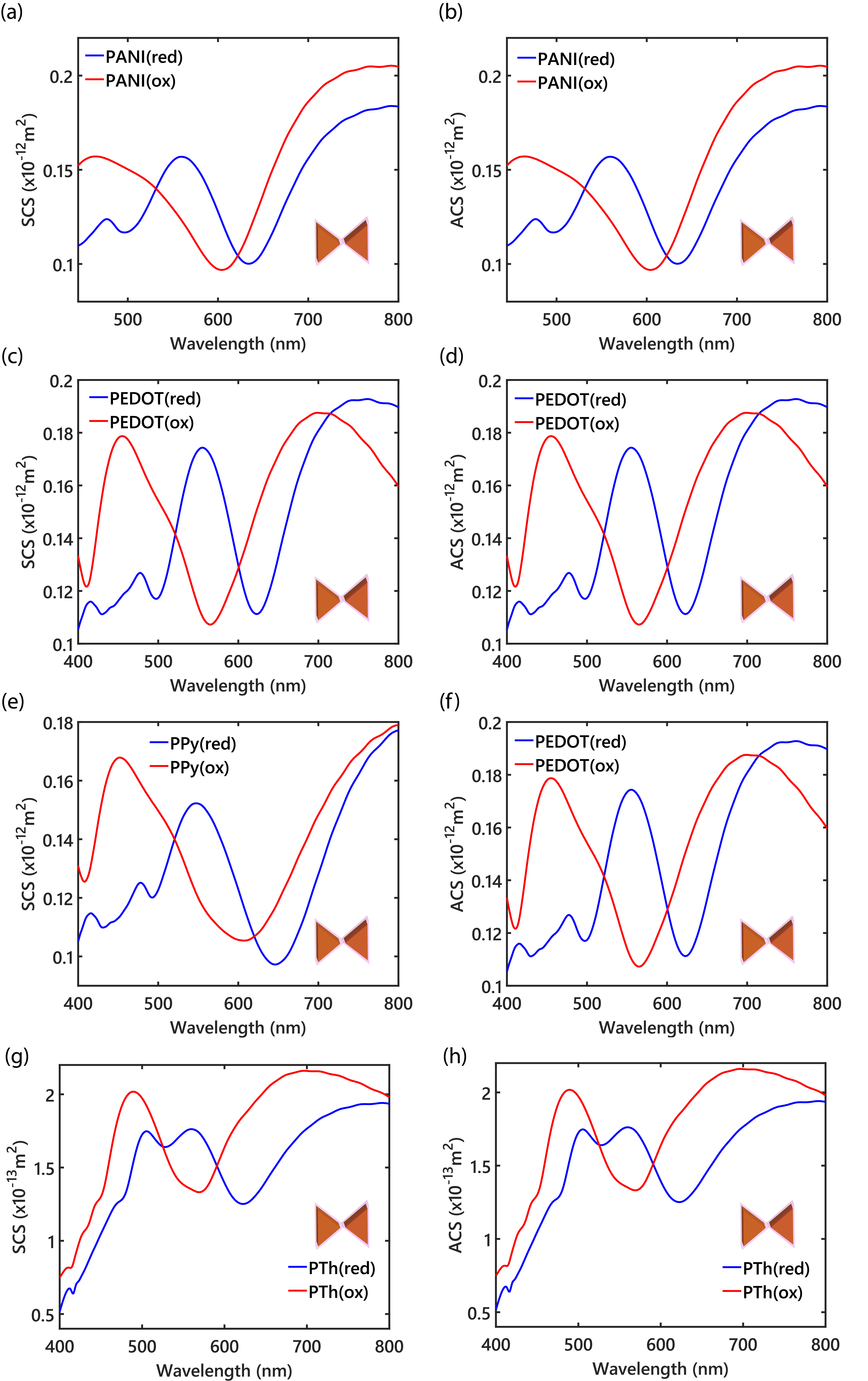}
\caption{(a, c, e, g) scattering cross-section (SCS) and (b, d, f, h) absorption cross-section (ACS) for bow tie shaped particle on Au mirror.}
\label{bow_tie_scs_acs}
\end{figure}

\begin{figure}
\centering
\includegraphics[width=0.5\textwidth]{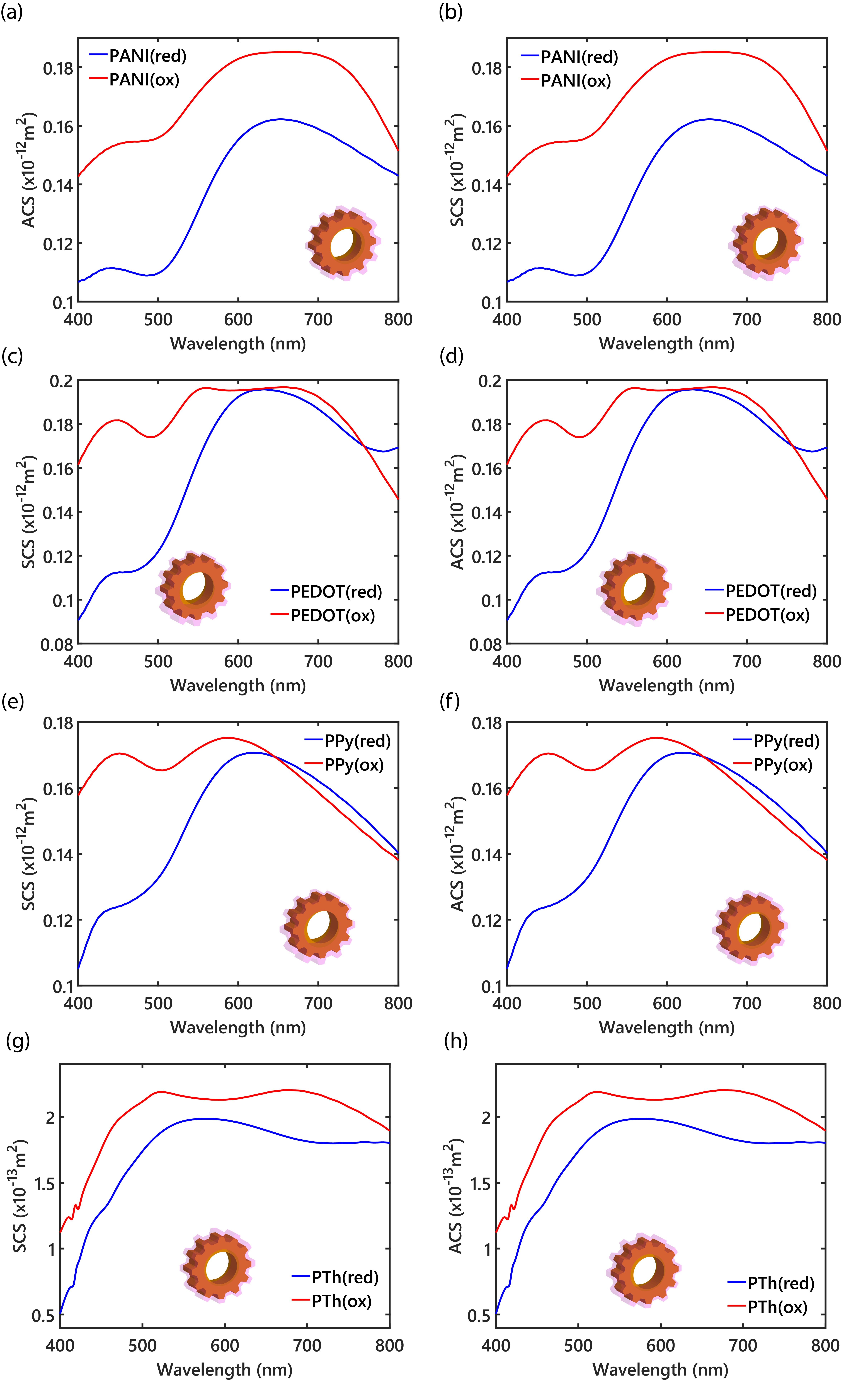}
\caption{(a, c, e, g) scattering cross-section (SCS) and (b, d, f, h) absorption cross-section (ACS) for a gear-shaped particle on Au mirror.}
\label{gear_shape_scs}
\end{figure}

\subsection{Impact of TiN upper mirror}
\begin{figure}
\centering
\includegraphics[width=0.4\textwidth]{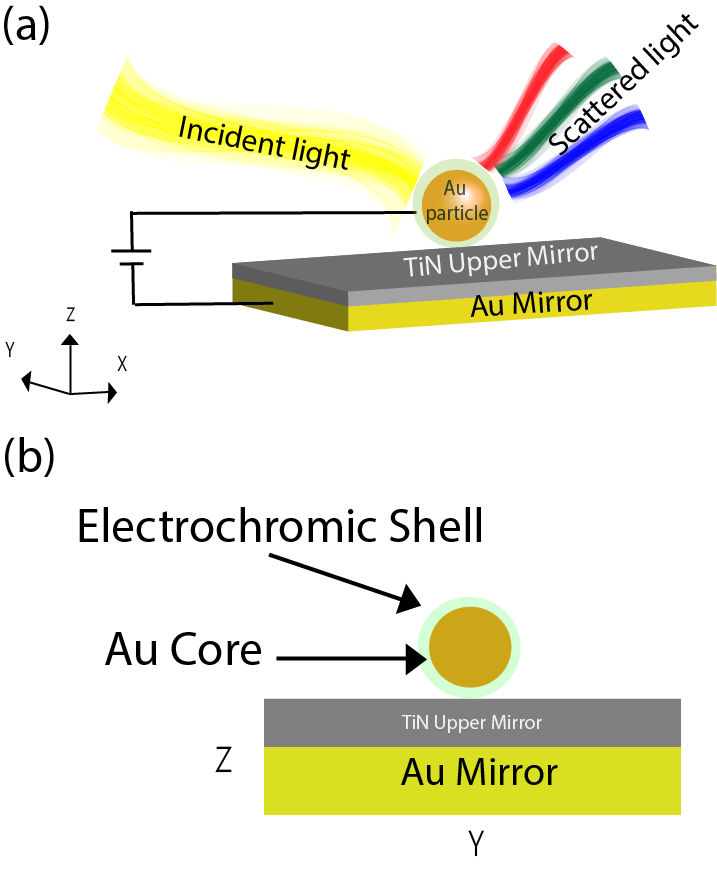}
\caption{Illustration of Au nanoparticle on TiN mirror structure over Au substrate-(a) perspective view, and (b) side view.}
\label{TiN_layer}
\end{figure}

From equation \ref{peak shift}, it is evident that
The change in resonant wavelength, $\Delta\lambda$ is inversely proportional to dielectric permittivity slope $d\epsilon_{m}/d\lambda$. Therefore, to increase the color tunability, it is needed to decrerase the permittivity gradient between shell and mirror. Hence, we introduced TiN, a semi-metallic alternative plasmonic material, as a spacer layer on the Au mirror. Such spacer layer allowed us to increase the tunability of nanopixel scattering profile and make it more color-dynamic. Previously reported promising plasmon tuning capability in smooth as well as rough TiN layer motivated us to choose it as the alternative plasmonic material on the upper layer\,\cite{Islam:23_Plasmon_Tuning_TiN}. By adding an additional layer in such NPoM design, the plasmonic field was tightly confined in the TiN layer. The LSPs of nanoparticle got coupled with their counter charges in Au mirror underneath the TiN upper layer. This ultimately resulted in confined plasmoic hotspot in TiN layer. This phenomenon can be compared with plasmonic hot-spot between nanodimer\,\cite{DiMartino2020_TiN_Layer}. The perspective and cross-sectional view of the described structure is shown in Fig.\,\ref{TiN_layer} where an additional 15nm TiN was overlaid on Au surface. The simulation results supported the theoretical assumptions in Fig.\,\ref{TiN_color_Switching}.  A more comprehensive range of color switching was possible than that of Au mirror.

In Table\,\ref{tab:CIELAB2000_color_difference}, the CIELAB2000 color difference through redox cycle in electrochromic materials studied are tabulated. The color difference values were determined utilizing CIELAB2000 color difference formula using MATLAB. Previously calculated CIE 1931 data was converted to recently developed CIELAB color space. In the case of PANI and PEDOT, the color difference gets higher along with incremental voltage on the TiN mirror whereas a slight decrement was observed in the case of PPy and PTh. 

\subsection{Shell thickness variation}
Shell thickness is a vital parameter of scattering modulation.  For the core-shell type structure, the net volume of the electrochromic layer can be determined by subtracting the core metal volume from the total nanoparticle volume. Geometric parameters of proposed nanoparticle structures are shown in Fig.\,\ref{index} and the estimated electrochromic shell layer volumes are tabulated in Table\,\ref{tab:shell_volume_formula}.

\begin{table}
    \centering
    \caption{Electrochromic shell volumes for three different structures studied.}
 \begin{tabular*}{21pc}{@{\extracolsep{\fill}}ll@{}}
 \hline
         Structure& Shell Volume\\
         \hline
         Sphere& $\frac{\mbox{4}}{\mbox{3}}\pi \left[\left(\mbox{r+t}\right)^3\mbox{-r}^3\right]$\\
         bow tie& $\left\{\frac{\mbox{1}}{\mbox{2}}\left(\mbox{b+t}\right)\left(\mbox{h+t}\right)\left(\mbox{l+t}\right)\right\}\mbox{-}\left(\frac{\mbox{1}}{\mbox{2}}\mbox{bht}\right)$\\
         Gear& \makecell{$\left[\left[\pi \left(\mbox{h+t}\right)\left\{\left(\mbox{r}_1\mbox{+t}\right)^2-\left(\mbox{r}_2\mbox{+t}\right)^2\right\}\right]
\mbox{-}\left\{\left(\mbox{j+t}\right)^3\mbox{n}\right\}\right]$\\
$\mbox{-}\left[\left\{\pi \:\mbox{h}\left(\mbox{r}_1^2-\mbox{r}_2^2\right)\right\}\mbox{-}\left\{\mbox{j}^3\:\mbox{n}\right\}\right]$}\\
         \hline
    \end{tabular*}
    
    \label{tab:shell_volume_formula}
\end{table}

\begin{figure}
\centering
\includegraphics[width=0.5\textwidth]{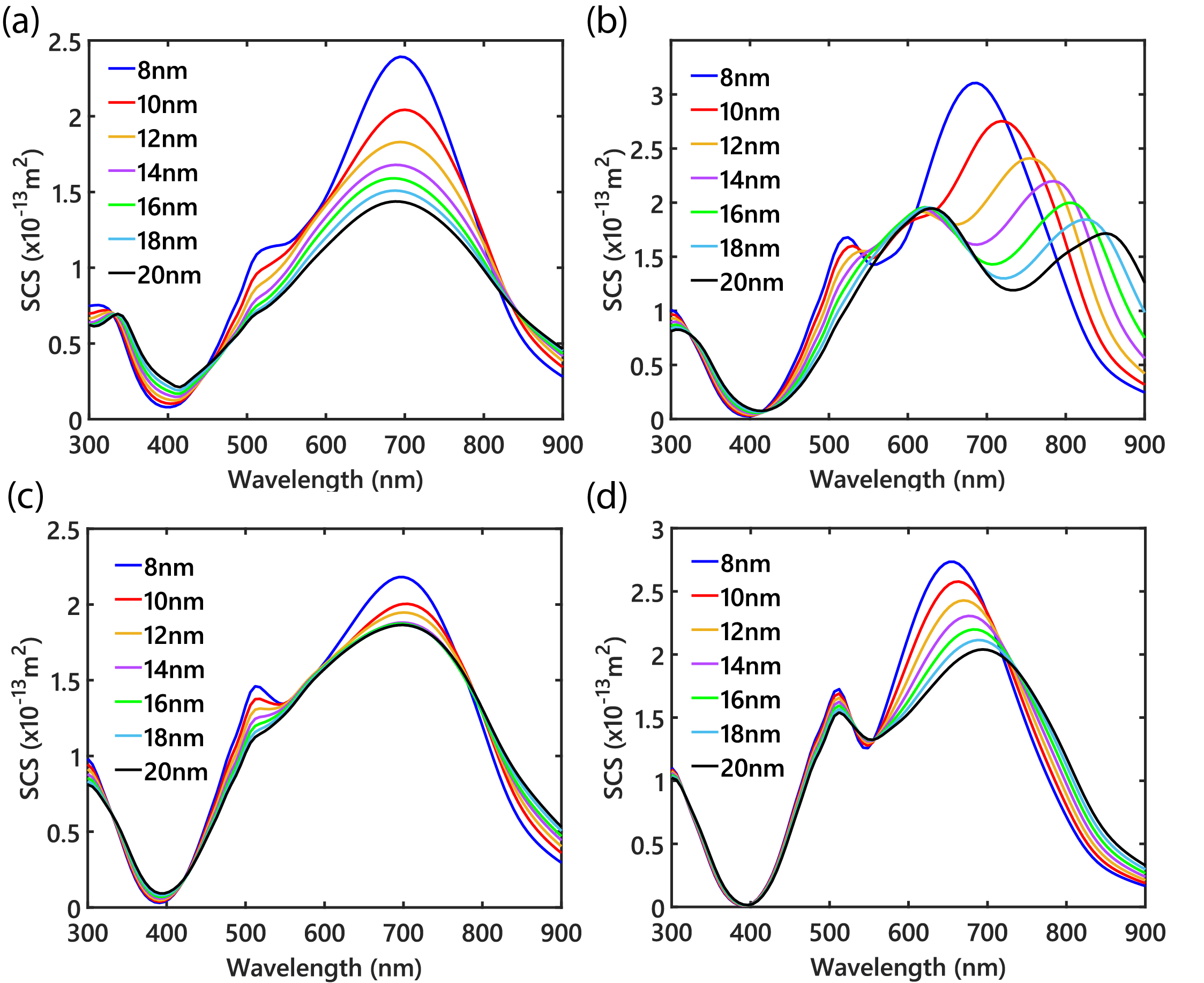}
\caption{Shell thickness-dependent SCS peak shift for (a) PANI, (b) PEDOT, (c) PPy, and (d) PTh nanosphere: shell thickness was varied from 08nm to 20nm}
\label{thickness_variation}
\end{figure}

The net volume of metallic nanoparticles can be varied in many ways. However, the expected substantial change in extinction spectrum gets maximized when the shape changes occur along the smallest cross-sections of the nanoparticle as this results in affecting the eigenvalues and their coupling weights to the electromagnetic field\cite{Sandu2012_shape_change_through_smallest_cross_section}.  On the other hand, the volume of electrochromic material can be changed via shell thickness variation. This is more effective in the device described in this investigation. Because we considered a faradaic reaction that resulted in a redox cycle in NPoM structures. This scheme's process is mostly shell-dominated\,\cite{Kim_2021_ec_metal_np}. 
From a fabrication perspective, the electrochromic layer thickness can be controlled by adjusting the monomer amount in the nanoparticle coating process. In our proposed core-shell nanoparticles, shell layer thickness is crucial for controlling optical resonance and electron dynamics in redox chemistry. The shell thickness not only determines the critical gap to the underlying mirror but also ensures sufficient spacing between neighboring Au NPs, reducing optical coupling and preventing aggregation. The gradual shift of shell thickness-dependent resonant wavelength is shown in Fig.\,\ref{thickness_variation}.

\begin{table}
\centering

\begin{tabular}{l l l l l} 
\hline
\multicolumn{5}{l}{RGB Components in redox cycle (in scale of 255)} \\
\hline
EC & Oxidization state & Red & Green & Blue \\
\hline
PANI & Reduced & 204 & 172 & 90 \\
 & Semi-Oxidized & 102 & 176 & 156 \\
 & Oxidized & 161 & 126 & 176 \\
PEDOT & Reduced & 174 & 201 & 41 \\
 & Semi-Oxidized & 25 & 178 & 167 \\
 & Oxidized & 100 & 141 & 188 \\
Ppy & Reduced & 175 & 180 & 108 \\
 & Semi-Oxidized & 83 & 165 & 172 \\
 & Oxidized & 147 & 142 & 171 \\
Pth & Reduced & 118 & 177 & 148 \\
 & Semi-Oxidized & 93 & 147 & 185 \\
 & Oxidized & 125 & 184 & 135 \\
\hline

\end{tabular}
\end{table}

\subsection{Shape dependency}
According to Mie theory's extension, there is a strong correlation between particle shape and LSPR frequency. Additionally, the plasmonic resonance shift is directly influenced by the permittivity profile, which is directly controlled by the carrier concentration. The plasmonic oscillation occurs at sharp edges of the nanostructure where heterogeneous dielectric index results in light-matter interaction. Free electrons in nanoparticles oscillate with incident light at LSPR wavelength. This effect can be explained by the Mie Theory for a metal sphere surrounded by a dielectric medium. Hence, the extinction cross-section, $\sigma$ can be expressed by the following simplified formula\,\cite{nehl2008shape_dependent_plasmon},

\begin{equation}
    \sigma=\frac{18{\pi}V{\epsilon}^{3/2}}{\lambda}
    \frac{\mathbf{\epsilon_{2}}} {(\epsilon_{1}+2{\epsilon_{m}})^2+\mathbf{\epsilon_{2}^2}},
\label{Extinction_Cross_Section}
\end{equation}

where $V$ is the nanoparticle volume, $\lambda$ is the wavelength of light, $\epsilon_{m}$ is the dielectric constant of the medium, and $\epsilon=\epsilon_{1}$ + $i\epsilon_{2}$ is the complex dielectric constant of the metal nanoparticle. The resonance will occur when $\epsilon_{1}=-2\epsilon_{m}$. It can seem from Eq.\,(\ref{Extinction_Cross_Section}) that the total volume of a nanoparticle is one of the determinants of plasmonic resonance wavelength and there is no dependency of nanoparticle shape on the resonant wavelength. However, experimental results\,\cite{nehl2008shape_dependent_plasmon} suggest a strong shape-dependent property of plasmon resonance frequency. In equation \ref{peak shift}, $\chi$ represents the nanoparticle's refractive index and shape factor. The value of $\chi$ differs in spheres, bow ties, and gear. Therefore, it was expected to obtain color shifts in different scales for different nanoparticle shapes. Until now, all the results described had a sphere as a nanoparticle, with the value of $\chi$= 2. The shape of plasmonic electrochromic nanoparticles significantly influences their localized surface plasmon resonances (LSPRs), which are crucial for their optical properties. Different shapes can lead to variations in the LSPR due to changes in the electromagnetic field coupling and the distribution of the electric field around the nanoparticles. In bulk materials, the mean free path of charge is significantly smaller compared to the dimensions of the system. However, the transport characteristics are governed by dimensions when the structure is comparable to or smaller than the mean free path of charge. The same applies to optical phenomena. The formulations for bulk material were adapted in nanostructures to consider the mesoscopic effect\,\cite{Mortensen_2021_mesoscopic}. Hence, we comprehensively analyzed the impact of nanostructure shapes, dimensions, and surface roughness during plasmonic structure modeling. Figure\ref{TiN_color_Switching}
and \ref{gear_shape_scs} demonstrate the scattering profile of bow tie and gear-shaped particles.

Unlike most studies focusing on spherical plasmonic nanoparticles, we investigate bow-tie and gear-shaped nanoparticles along with the sheperical ones, showing that shape of nanoparticle core and shell influences plasmonic resonance tuning. This adds a new dimension to the design flexibility of electrochromic plasmonic displays.
In these results, we determined the direct involvement of shape and particle volume in the interaction of incident light on the nanoparticles.  Localized surface plasmons (LSPs) got excited in sub-wavelength scale metallic nanostructures, where light is confined to nanoparticle surfaces. Plasmonic nanoparticles, such as sphere, bow tie, and gear, studied here facilitated the excitation of free waves. The properties of LSPs in those structures are influenced by geometrical factors such as size, shape, shell material composition, and proximity to other plasmonic structures, which offered significant flexibility for light manipulation.

LSP resonances can be calculated analytically for the studied nanoparticle shapes using the quasi-static approximation. In this approximation, the nanoparticles behave like oscillating dipoles characterized by their polarizability, denoted as $\alpha$.

\begin{equation}
    \alpha=V
    \frac{\mathbf{\epsilon_{m}-\epsilon_{d}} }{{\epsilon_{m}}+2\mathbf{\epsilon_{d}}}.
\label{Polarizability}
\end{equation}

Where $\epsilon_{m}$ and $\epsilon_{d}$ are the permittivity of the core and shell material, and V is the volume of the nanoparticle. In case of sphere, V=$4\pi R^3$, for bow tie, V= $bht$ (b, h, and t are the base, height and thickness of bow tie triangle), and in case of gear, the volume is, V=$\pi h (r_{1}^{2}-r_{1}^{2})+j^{3} n$($r_1$ and $r_2$ are the outer and inner radius of gear; h, j, and n represents the thickness, teeth dimension and number of teeth respectively).

There is square-proportional relationship between polarizability and scattering.  Theoretically, the scattering cross section from any nano-particle, whose size is comparable to the wavelength of light can be expressed as,

\begin{equation}
    C_{scs}=\frac{8\pi}{3}k^{4}a^{6}\left| \frac{\epsilon-\epsilon_{m}}{\epsilon+2\epsilon_{m}} \right|^{2}.
    \label{scs}
\end{equation}

Similarly, absorption cross-section can also be expected by quasi-static approximation,

\begin{equation}
    C_{acs}=4\pi ka^{3}\left| \frac{\epsilon-\epsilon_{m}}{\epsilon+2\epsilon_{m}} \right|.
    \label{acs}
\end{equation}
The sum of these two cross-section quantities is known as 'extinction cross section', which implies the ratio of the total amount of light that gets escaped or absorbed by the nanoparticle to the total incident light.

\begin{equation}
    C_{ecs}=C_{scs}+C_{abs}.
    \label{ecs}
\end{equation}

The important observation from the above-mentioned formulae is the dependency of the scattering cross section on particle radius is much higher than the scattering cross section. $C_{scs}$ scales as $a^{6}$ where $C_{acs}$ scales as $a^{3}$. So, absorption is dominant in nanoparticles with smaller radii, whereas scattering is dominant for the larger ones. Numerous research works indicate that even small alterations in shape can result in substantial changes in the extinction spectrum of metallic nanoparticles. Our study also reflects on it. 

In Figure \ref{gear_shape_scs}, the scs and acs profile of gear shape structures are shown. Our calculated result suggested that bigger nanoparticles such as gear demonstrated higher scattering (nearly flat scs curve), whereas the bow tie structures have some specific peaks in the scs profile, which indicated a lower average scattering cross section. This implies the theoretical assumption as explained in Equation\ref{scs}.  Another important observation from this result is that nanoparticles with larger aspect ratios and small volumes exhibited a higher shift in resonance frequency. For instance, we can consider the scattering peak shifting property of PEDOT shells. When this electrochromic material was imposed on a bow tie-shaped nanoparticle, it exhibited more than 100nm of a shift in resonance wavelength compared to its lower tunability in the case of the gear-shaped nanoparticle variant. The reason behind this is the higher area-volume ratio of bow tie structure compared to gear. This significant shift allowed the development of a specific light emitter within the RGB spectra. These shape-dependent properties are essential for designing plasmonic nanoparticles with desired optical characteristics for sensing, imaging, and optoelectronics applications.

In equation \ref{peak shift}, $\chi$ represents the nanoparticle's refractive index and shape factor. The value of $\chi$ differs in spheres, bow ties, and gear. Therefore, it was expected to obtain color shifts in different scales for different nanoparticle shapes. Until now, all the results described had a sphere as a nanoparticle, with the value of $\chi$= 2. The local field enhancement described in the previous subsection is subject to further improvisation by shape variation and employing nanostructures such as optical nanoantennas, slanted bound states\,\cite{Hsu_21_OE_local_field_enhancement}. The shape of plasmonic electrochromic nanoparticles significantly influences LSPRs, which are crucial for their optical properties. Different shapes can lead to variations in the LSPR due to changes in the electromagnetic field coupling and the distribution of the electric field around the nanoparticles. Sarker \textit{et al.} demonstrated structural colorization contrast on ingenious Si–SiO$_2$–Si dielectric metasurface\,\cite{soikot_metasurface_color}. Figs.\,\ref{TiN_color_Switching} and \ref{gear_shape_scs} demonstrate the scattering profile of bow tie and gear-shaped particles. Numerous research works indicated that even small alterations in shape can result in substantial changes in the extinction spectrum of metallic nanoparticles. The presence of grooves in EC device structure has several advantages, such as an increased surface area and enhanced ion transport capabilities\,\cite{Hao_2023_groove}.  Our study reflects on the effect of nanoparticles' size and aspect ratio on resonant frequency shift through the redox cycle. Nanoparticles with a larger aspect ratio and small volume exhibited a higher shift in resonance frequency. For instance, PEDOT shells' scattering peak shifting property increases when electrochromic material was imposed on a bow tie-shaped nanoparticles. It exhibited more than 100nm of a shift in resonance wavelength compared to its lower tunability in the case of the gear nanoparticle variant. This shift allowed the development of a specific light emitter within the RGB spectra. These shape-dependent properties are essential for designing plasmonic nanoparticles with desired optical characteristics for sensing, imaging, and optoelectronics applications.

\subsection{Inorganic and organic electrochromic material}

\begin{figure}[h]
\centering
\includegraphics[width=0.5\textwidth]{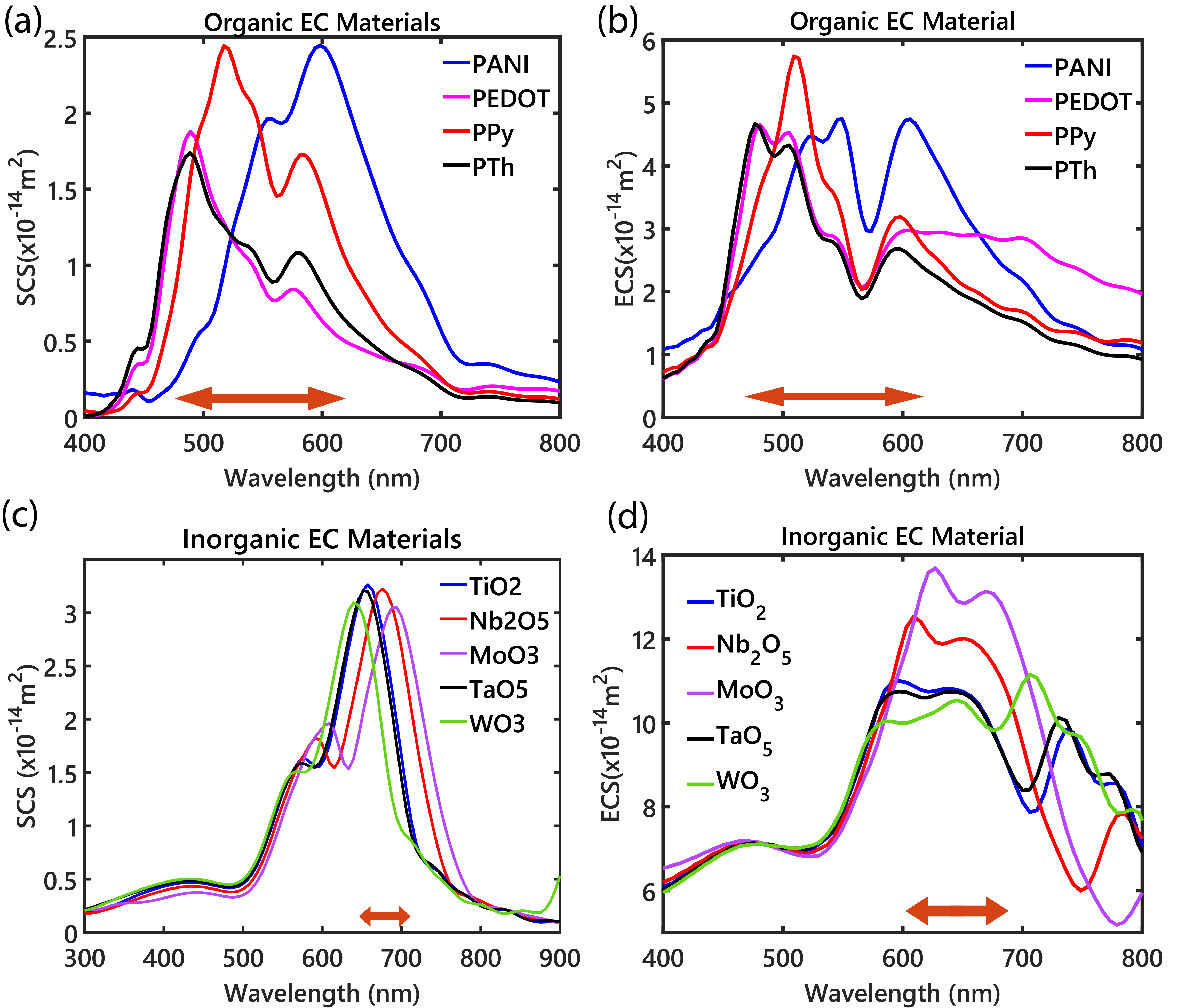}
\hspace{-50pt}
\caption{Scattering and extinction cross section profile variation of different organic electrochromic materials: (a) and (b) demonstrates dynamic range of SCS and ECS for organic electrochromic materials, (c) and  (d) denotes narrow range of SCS and ECS profile of inorganic electrochromic materials.}
\hspace{-50pt}
\label{comparison_organic_inorganic_inorganic}
\end{figure}

Electrochromic materials, the primary component of electrochromic devices, have various optical characteristics that lead to their use in numerous applications. One of the key factors affecting the redox cycle tolerance of electrochromic devices is the choice of electrolyte material. Two categories of electrochromic materials exist: inorganic and organic. We conducted a comparative analysis and performance evaluation of different electrochromic materials.

Hexacynometallates and group IV-VI transition-metal oxides are popular among inorganic materials. In those inorganic materials, electron delocalization between the mixed-valence states occurs, resulting in different light absorption and material color changes\,\cite{Patel2017all}.  Li$^{+}$ and Na$^{+}$ are two examples of monovalent ion solutions whose redox peaks were significantly lower in K$^{+}$ than in thin Cu$_2$O electrochromic film\,\cite{RISTOVA2007_electrolyte}. 
The majority of the inorganic electrochromic materials are transient metal oxides such as vanadium, iridium, molybdenum, tungsten, and niobium oxides. According to Cong's report on WO$_3$ \cite{cong2016tungsten}, these oxides create the MO$_6$ octahedra structure, in which M is a transitory metal. These layered structures enable ion transport through conduits, chains, or interstitial sites. On the other hand, Granqvist \textit{et al.} showed that those materials' empty d bands become occupied by the cathodic charge injection and can be used to undergo a new intra-band transition that results in color change \cite{granqvist2014electrochromics}. Among the transitory metal oxides mentioned above, tungsten oxide is the most effective and exhibits excellent color change. It displays a blue condition after the reaction shown below\,\cite{zhang2020nanostructured}.

   \begin{equation}
 \mbox{WO}_3\mbox{(bleached)} + \mbox{xe}^- + \mbox{xM}^+ \longleftrightarrow  \mbox{M}_x\mbox{WO}_3\mbox{(colored)}
\end{equation}

On the other hand, organic electrochromic films demonstrate several colored states, excellent visual contrast, improved switching speeds, and good staining efficiency\cite{monk1999viologens}. Most often used organic electrochromic compounds are viologens (1,1'-disubstituted-4,4'-bipyridinium salts)\cite{shah2019viologen}. Viyogens exhibit electrochromism in three different redox states: radical cation $V^+$, di-cation $V^{2+}$, and neutral form V, which result in three distinct chromisms\,\cite{zhang2020nanostructured}. However, organic electrochromic material suffers from low stability and oxidation number\,\cite{mortimer1999organic}. Adding inorganic electrochromic material gives the electrochromic conductive polymer better properties. The inorganic material also enhances the coloration efficiency and the switching speed, but it does not change the electrochemical behavior of the conductive polymer. Electrochromic films such as WO$_3$, Nb$_2$O$_5$, and NiO are preferred for their high durability and stability,\cite{shchegolkov2021brief}. In the sense of color-changing speed, organic electrochromic materials are better than inorganic ones. The chemical stability of organic polymers can be attributed to this. However, this stability raises issues in the reversibility of electrochromic devices. Moreover, organic electrochromic materials are more color efficient (100 Sm$^{2}$C$^{-1}$ for PEDOT] and Lower efficiency [Ex: $\approx$ 40 Sm$^{2}$C$^{-1}$ for WO$_3$ and NiO). The easy flexibility and low production cost add plus points to organic polymers for wearable electrochromic devices. A summary table of our comparison between organic and inorganic electrochromic material is presented in Table. S1 of Supplementary Material.

\begin{table}
    \centering
    \caption{CIELAB2000 color difference through redox cycle of electrochromic materials used in the eNPoM nanopixel}
    \begin{tabular}{ccc}
    \hline
         Material& \shortstack{Voltage Applied on\\ TiN Mirror (V)}& CIELAB2000 difference
\\
\hline
         PANI&  2& 8.85
\\
         &  2.5& 8.50
\\
         &  5& 12.21
\\
         PEDOT&  2& 26.50
\\
         &  2.5& 26.34
\\
 & 5&54.55
\\
 Ppy& 2&38.59
\\
 & 2.5&41.27
\\
         &  5& 33.53
\\
         PTh&  2& 38.59
\\
         &  2.5& 38.89
\\
         &  5& 31.77
\\
\hline
    \end{tabular}
    \label{tab:CIELAB2000_color_difference}
\end{table}

\section{Comparative analysis}

Recent advancements in electrochromic plasmonic nanopixels, particularly in exploiting nanophotonic and nanoplasmonic effects, have made these promising in diverse applications. Numerous research groups are working on advanced applications in full-color displays, enhancing transmission, and obtaining ultra-high on-off ratios and sensing abilities. Our study is based on four electrochromic polymers: PANI, PEDOT, PPy, and PTh. Other less common electrochromic polymers show wavelength shifts in the visible regime. In this context, poly(3,4-propylenedioxythiophene)  (PProDOT) can be a candidate \,\cite{Ledin_2021_Au/PProDOT}. A recent study reported methodical techniques for developing hybrid polymer-metal nanomaterials that are plasmonically active and also have electrochemically adjustable localized surface plasmon resonances (LSPRs)~\cite{Ledin_2021_Au/PProDOT}. Ledin \textit{et al.} proposed ultra-thin plasmonic electrochromic hybrid films composed of gold nanorods and EC layers\,\cite{ledin_2021_ec_nanorod}. When an electrical potential was applied, this scheme demonstrated significant, consistent, and reversible LSPR modulation ranging from 25 to 30nm. Peng \textit{et al.}  developed an NPoM-based electrochromic pixel\,\cite{peng_2019_nanopixel}. In that study, they presented a scalable method for creating electrically driven color-changing metasurfaces by combining plasmonic metasurfaces with electrochromic materials.  Moreover, these pixels were electrically tunable across more than 100nm wavelength ranges. Leroux \textit{et al.} reported on new active molecular plasmonic devices where the electrochemical switching of a nanometric film of conductive polymer between its reduced and oxidized states utilizing a gold disk array. As the electrochromic layer, they used PANI and obtained a wavelength shift of 63nm\,\cite{Leroux_Au_PANI}. Jiang \textit{et al.} reported nanorod-based structures served as dynamic plasmonic switches. These individual core/shell nanostructures demonstrate impressive switching capabilities, achieving a modulation depth of 10 dB and a scattering peak shift of 100nm. Additionally, these nanostructures were assembled on substrates to create large-scale monolayers that exhibited significant collective plasmonic switching behavior. However, the switching time of their proposed switch was more than 100 s\,\cite{Jiang_Au_PANI}, which was an obstacle in incorporating such structures in display devices.
\\
WO${_3}$ is the most successful inorganic electrochromic material. Many research works reported the possibilities of utilizing WO${_3}$ in display devices. Wang and Li \textit{et al.} obtained a peak resonance of 498 and 620nm for colloidal sol and rectangular array structures, respectively. However, their wavelength shifting capabilities (31 and 55nm, respectively) were much lower than that of the organic polymers\,\cite{Wang_Ag_WO3, Li_Al_WO3}. This reflects the analysis comparing organic and inorganic materials in the previous section. Alongside the electrochromic layer-based approaches, single sphere and dimer-based Au nanoparticles were proposed for the detection of reversible electrochemical potential-driven anion adsorption by developing single-particle plasmon voltammetry (spPV) \,\cite{Byers2016-Au}. In our analysis, we found PEDOT and PTh to be the most tunable electrochromic polymers for all shapes of nanoparticles. PEDOT films could dynamically change the scattering wavelength by more than 40nm. keeping the geometry the same, PEDOT films can be successfully imposed on RGB tunable pixels. On the other hand, the PTh shell demonstrated the possibility of tuning more than 50nm wavelength. Stockhausen \textit{et al.} reported a similar significant shift ($\approx$180) in the localized surface plasmon resonance (LSPR) of gold nanoparticle arrays when covered by PEDOT\,\cite{Stockhausen_Au_PEDOT}. Our study revealed the possibility of PEDOT and PTh film utilization in widespread application. A comparative report of similar organic polymer-based electrochromic material-based devices with the studied ones are tabulated in Table\,\ref{tab:comparison}.

\begin{table*}
    \centering
     \caption{Comparison of wavelength shift and color switching capacity among our work and recently reported eNPoM structures}
{    \begin{tabular}{cccccc}
    \hline
         Materials&  Structure&  Peak wavelength (nm)&  Wavelength shift, $\delta\lambda$ (nm)&  Switching time (s)& Reference
\\
\hline
 Au/PANI& NPoM& 618& -105& -&This Work\\
         Au/PANI&  Disk array&  631&  -62&  $<$10& \cite{Leroux_Au_PANI}\\
 Au/PANI& NPoM& 680& -100& 0.02&\cite{peng_2019_nanopixel}\\
 Au/PEDOT& Nanosphere& 602& -44& $<$0.3&This Work\\
         Au/PEDOT&  Disk array&  720&  +180&  -& \cite{Stockhausen_Au_PEDOT}
\\
         Ag/WO$_{3}$&  Colloidal sol&  498&  -31&  $<$5& \cite{Wang_Ag_WO3}
\\
         Au/PANI& Nanorods&  740&  -100&  $<$240& \cite{Jiang_Au_PANI}
\\
         Al/WO$_{3}$& Rectangle array&  620&  -55&  $<$20& \cite{Li_Al_WO3}
\\
         Au/PPy&  NPoM&  630&  -73&  -& This Work\\
         Au&  Dimer&  621.5&  +16&  -& \cite{Byers2016-Au}\\
         Au/PProDOT&  EC immersed Nanorod&  700&  +25/+30&  -& \cite{Ledin_2021_Au/PProDOT}\\
         Au/PTh&  NPoM&  488&  -44& -& This Work\\
         \hline
    \end{tabular}
   }
    \label{tab:comparison}
\end{table*}

\section{Proposed fabrication method} 

Several high-precision fabrication schemes such as electron beam lithography (EBL)\cite{martin2014_Al_fabrication}, focused ion beam (FIB) milling\cite{Chen_2016_FIB_Milling_acsnano} are available to fabricate metallic nanoparticle precisely. A possible fabrication scheme for the proposed nanopixel is to proceed to the bottom-up technique. The core metal should be homogeneous to fabricate the proposed core-shell structure. Dispersing the core material (Au) in an alcohol medium using ultrasonication can effectively ensure uniform dispersion. Afterward, the metal can be centrifuged to make nano-spheres. Proper surface treatment is needed to enhance the adhesion between metal and electrochromic material, as the sphere surface will be coated with electrochromic materials. Coupling agents such as saline can be utilized to introduce some functional groups on the metal surface, which will enhance the adhesion where electrostatic forces cause the EC monomer to coat Au NPs, ultimately forming a seed layer for further polymer growth. Alongside the core material, electrochromic material treatment is also vital in electrochromic nanopixel fabrication. Proper solvent selection for different electrochromic polymers is important to make a solution of electrochromic material. Chemical vapor deposition (CVD) or electrodeposition can be used to coat the Au cores with an electrochromic shell. Colloidal Au NPs can be coated with thin, continuous electrochromic shells using surfactant-assisted chemical oxidative polymerization. After the deposition, an organic stabilizer such as Polyvinylpyrrolidone (PVP) is added to avoid unexpected agglomeration of the coated nanopixel. Proper incubation with acidic ammonium persulfate (APS) can be done to speed up the polymerization of the coating. As with other nanoparticles, annealing can improve the crystallinity and electrochromic properties of the core-shell structure. Finally, the nanopixels can be washed and re-dispersed in SDS solution. For easy implementation of frequent voltage deviation between nanoparticles and mirrors, ion gel formation can be a possible way. It will significantly stabilize the Au nanoparticles, enabling them to quickly and reversibly adjust the carrier concentration when the external voltage is altered. The proposed bi-layer mirror can be grown on alkaline earth metal oxide substrates utilizing pulse laser deposition. An implementation of such bi-layer film was demonstrated by Martino \textit{et al.} \cite{DiMartino2020_TiN_Layer}.

\section{Conclusion}
We reported four electrochromic conducting polymers here and evaluated their coloration capability as nanopixels. We calculated the optical properties of these conducting organic polymers at different oxidation states through atomistic simulations. In the case of the PANI chain, our study revealed a refractive index peak of around 1650nm and an extinction coefficient of around 850nm. PANI, PEDOT, and PPy exhibit two prominent peaks in their n-k spectra for the reduced, semi-oxidized, and oxidized states. The intensity of the peaks varies between the reduced and oxidized states. Except for PANI, the oxidized states show higher peak intensities than the reduced states in all electrochromic materials, resulting in higher absorption. We designed eNPoM nanopixels containing spherical, bow tie, and gear-shaped core-shell structures to implement low-power-consuming nanoscale pixels. We used PANI, PEDOT, PPy, and PTh shells to achieve wavelength tunability of 10nm, 20nm, 8nm and over 80nm, respectively. Furthermore, a bow tie-shaped Au structure coated with PEDOT resulted in an absorption peak shift exceeding 100nm. The electric field enhancement at shell-mirror and shell-air junction was calculated and analyzed. Moreover, the volumetric calculation of electrochromic shell material was performed to focus on the effect of shell thickness. To represent the color-changing capability of different electrochromic shells, we determined the chromaticity coordinates, represented by x and y in CIE 1931 color space, and the diagonal shift during the redox cycle. From the perspective of the present display manufacturing industry, we included a detailed comparative analysis between organic and inorganic materials based on recent research findings and our calculated results.
Our study on the proposed innovative electrochromic plasmonic nanopixel includes the integration of theoretical modeling and simulations, and the exploration of tunability through nanoparticle shape, shell thickness, and external stimuli. The combination of DFT, FDTD, and experimental feasibility studies makes this research both comprehensive and pioneering. Moreover. The proposed structures can attain easy color switching, wide-range color reproducibility, flexibility, and cost margin. Our study includes first-principles atomistic calculations of permittivity for electrochromic materials, which is rarely explored in existing literature. Moreover, in this study, we investigated bow-tie and gear shaped nanoparticle in electrochromic application for the first time. Inclusion of TiN as a mirror layer and multiple core-shell structures resulted in significant wavelength shift (20–80nm) for different electrochromic materials, allowing dynamic color tuning in nanopixels. This comprehensive study will give an insight into fast-switching, highly tunable next-generation nanopixel display design utilizing organic electrochromic materials. Our proposed structure can be used in regular display panels, interactive boards, road sign displays, and encryption systems.

\appendix

\printcredits

\bibliographystyle{model1-num-names}

\bibliography{cas-refs}


\bio{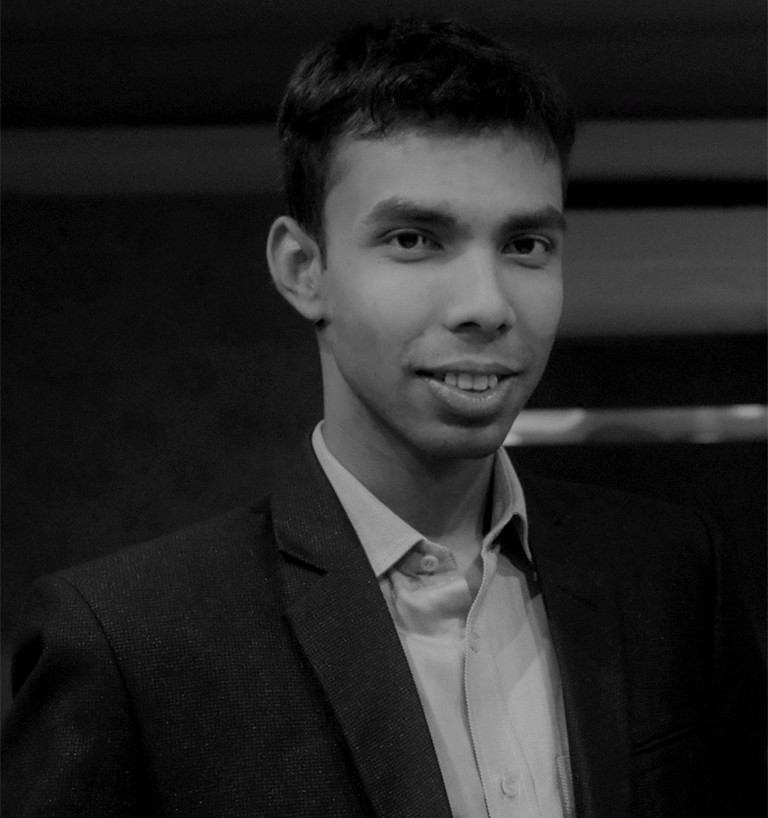}
Md. Shariful Islam has graduated from Shahjalal University of Science and Technology, Sylhet, Bangladesh on electrical and electronic engineering. Now, completing MSc. at Bangladesh University of Engineering and Technology, Dhaka, Bangladesh. As a profession, he is working in Bangladesh Telecommunications Company Limited as an assistant manager. He has also completed various certificate courses, including Industrial Technology on Electrical Engineering and Instrumentation, Professional Networking, Teaching-Learning Process, and other certifications. His research interests lie in the areas of photonics, plasmonics, digital electronics, VLSI and optoelectronic devices. Md. Shariful Islam have interest in exploring both the theoretical and experimental aspects of devices, which include design, experimental realization and characterization of novel electronic, optoelectronic devices.

\endbio

\bio{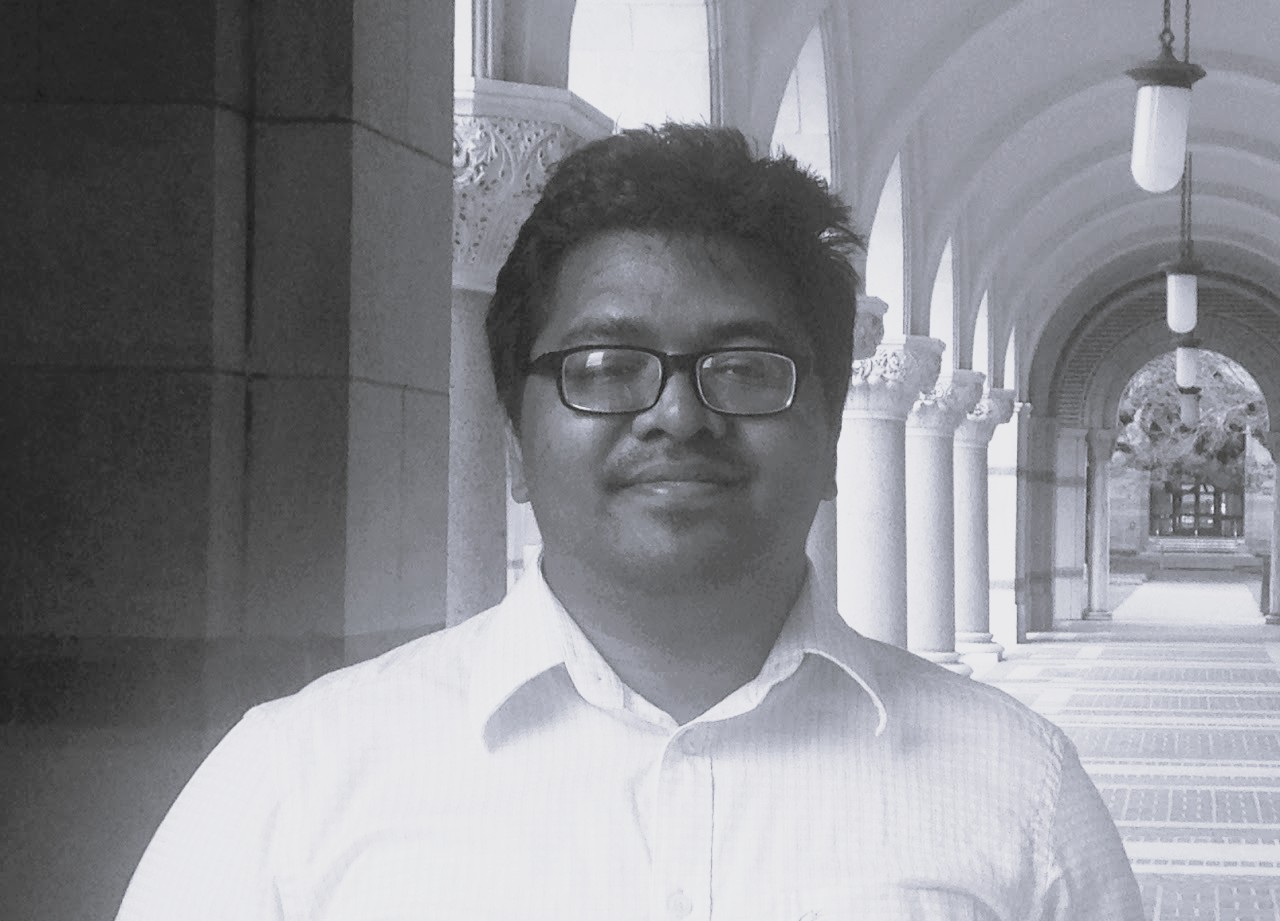}
Dr. Ahmed Zubair is a Professor in the Department of Electrical and Electronic Engineering (EEE) at Bangladesh University of Engineering and Technology (BUET). He completed his B.S. and M.S. degrees in Electrical and Electronic Engineering from BUET in 2009 and 2011, respectively, and earned his Ph.D. in Electrical and Computer Engineering from Rice University, USA, in 2017. Dr. Zubair's research interests encompass nanotechnology, nanomaterials, wearable optoelectronics, terahertz technology, electronic and photonic device fabrication and characterization, renewable energy, and ultrafast optical phenomena. He leads the Nanoscale Simulation, Characterization, and Fabrication (NSCF) Lab at BUET, focusing on theoretical investigations and practical applications in these areas. As an active member of the scientific community, Dr. Zubair holds senior memberships in several professional organizations, including the Institute of Electrical and Electronics Engineers (IEEE), Optical Society (OSA), American Association for the Advancement of Science (AAAS), American Physical Society (APS), and Materials Research Society (MRS). He serves as the Vice Chair of the IEEE Electron Devices and Solid-State Circuits Society (ED/SSCS) Bangladesh Chapter and as the Vice President of the OSA Bangladesh Section. Dr. Zubair has an extensive publication record, with numerous articles in peer-reviewed journals and conference proceedings. His work has contributed significantly to advancements in optoelectronics and photonics.

\endbio

\end{document}